\documentclass[fleqn, usenatbib]{mnras}

\usepackage{newtxtext,newtxmath}

\usepackage[T1]{fontenc}
\usepackage[utf8]{inputenc}
\usepackage[british]{babel}

\usepackage{amsmath}
\usepackage{mathtools}
\usepackage{graphicx}
\usepackage{xcolor}
\usepackage{xparse}
\usepackage[olditem, oldenum]{paralist}
\usepackage{mleftright}
\usepackage{xfrac}
\usepackage{microtype}
\usepackage{newunicodechar}

\usepackage{silence}
\ActivateWarningFilters[pdflatex-unicode-math]

\newunicodechar{†}{\dag}
\newunicodechar{‡}{\ddag}
\newunicodechar{…}{\ldots}
\newunicodechar{⋯}{\cdots}
\newunicodechar{⋮}{\vdots}
\newunicodechar{⋱}{\ddots}
\newunicodechar{⋰}{\iddots}
\newunicodechar{α}{\alpha}
\newunicodechar{β}{\beta}
\newunicodechar{γ}{\gamma}
\newunicodechar{δ}{\delta}
\newunicodechar{ε}{\varepsilon}
\newunicodechar{ϵ}{\epsilon}
\newunicodechar{ζ}{\zeta}
\newunicodechar{η}{\eta}
\newunicodechar{θ}{\theta}
\newunicodechar{ϑ}{\vartheta}
\newunicodechar{ι}{\iota}
\newunicodechar{κ}{\kappa}
\newunicodechar{ϰ}{\varkappa}
\newunicodechar{λ}{\lambda}
\newunicodechar{μ}{\mu}
\newunicodechar{ν}{\nu}
\newunicodechar{ξ}{\xi}
\newunicodechar{ο}{o}
\newunicodechar{π}{\pi}
\newunicodechar{ρ}{\rho}
\newunicodechar{ϱ}{\varrho}
\newunicodechar{σ}{\sigma}
\newunicodechar{τ}{\tau}
\newunicodechar{υ}{\upsilon}
\newunicodechar{φ}{\varphi}
\newunicodechar{ϕ}{\phi}
\newunicodechar{χ}{\chi}
\newunicodechar{ψ}{\psi}
\newunicodechar{ω}{\omega}
\newunicodechar{Α}{\mathrm{A}}
\newunicodechar{Β}{\mathrm{B}}
\newunicodechar{Γ}{\Gamma}
\newunicodechar{Δ}{\Delta}
\newunicodechar{Ε}{\mathrm{E}}
\newunicodechar{Ζ}{\mathrm{Z}}
\newunicodechar{Η}{\mathrm{H}}
\newunicodechar{Θ}{\Theta}
\newunicodechar{Ι}{\mathrm{I}}
\newunicodechar{Κ}{\mathrm{K}}
\newunicodechar{Λ}{\Lambda}
\newunicodechar{Μ}{\mathrm{M}}
\newunicodechar{Ν}{\mathrm{N}}
\newunicodechar{Ξ}{\Xi}
\newunicodechar{Ο}{\mathrm{O}}
\newunicodechar{Π}{\Pi}
\newunicodechar{Ρ}{\mathrm{P}}
\newunicodechar{Σ}{\Sigma}
\newunicodechar{Τ}{\mathrm{T}}
\newunicodechar{Υ}{\Upsilon}
\newunicodechar{Φ}{\Phi}
\newunicodechar{Χ}{\Chi}
\newunicodechar{Ψ}{\Psi}
\newunicodechar{Ω}{\Omega}
\newunicodechar{∑}{\sum}
\newunicodechar{∫}{\int}
\newunicodechar{∬}{\iint}
\newunicodechar{∭}{\iiint}
\newunicodechar{⨌}{\iiiint}
\newunicodechar{∮}{\oint}
\newunicodechar{∯}{\oiint}
\newunicodechar{∰}{\oiiint}
\newunicodechar{∇}{\nabla}
\newunicodechar{∂}{\partial}
\newunicodechar{√}{\sqrt}
\newunicodechar{∈}{\in}
\newunicodechar{∋}{\ni}
\newunicodechar{∉}{\notin}
\newunicodechar{∀}{\forall}
\newunicodechar{∃}{\exists}
\newunicodechar{∄}{\nexists}
\newunicodechar{∴}{\therefore}
\newunicodechar{∵}{\because}
\newunicodechar{⟨}{\langle}
\newunicodechar{⟩}{\rangle}
\newunicodechar{〈}{\langle}
\newunicodechar{〉}{\rangle}
\newunicodechar{⌊}{\lfloor}
\newunicodechar{⌋}{\rfloor}
\newunicodechar{⌈}{\lceil}
\newunicodechar{⌉}{\rceil}
\newunicodechar{∼}{\sim}
\newunicodechar{∝}{\propto}
\newunicodechar{∞}{\infty}
\newunicodechar{ℵ}{\aleph}
\newunicodechar{ℏ}{\hbar}
\newunicodechar{℘}{\wp}
\newunicodechar{ℓ}{\ell}
\newunicodechar{∅}{\emptyset}
\newunicodechar{×}{\times}
\newunicodechar{⋅}{\cdot}
\newunicodechar{÷}{\div}
\newunicodechar{⋆}{\star}
\newunicodechar{∘}{\circ}
\newunicodechar{⋄}{\diamond}
\newunicodechar{⊕}{\oplus}
\newunicodechar{⊖}{\ominus}
\newunicodechar{⊗}{\otimes}
\newunicodechar{⊘}{\oslash}
\newunicodechar{⊙}{\odot}
\newunicodechar{±}{\pm}
\newunicodechar{∓}{\mp}
\newunicodechar{≈}{\approx}
\newunicodechar{≃}{\simeq}
\newunicodechar{≡}{\equiv}
\newunicodechar{≠}{\ne}
\newunicodechar{≥}{\ge}
\newunicodechar{≤}{\le}
\newunicodechar{≳}{\gtrsim}
\newunicodechar{≲}{\lesssim}
\newunicodechar{≫}{\gg}
\newunicodechar{≪}{\ll}
\newunicodechar{⊂}{\subset}
\newunicodechar{⊃}{\supset}
\newunicodechar{⊆}{\subseteq}
\newunicodechar{⊇}{\supseteq}
\newunicodechar{⊈}{\nsubseteq}
\newunicodechar{⊉}{\nsupseteq}
\newunicodechar{≔}{\coloneqq}
\newunicodechar{≕}{\eqqcolon}
\newunicodechar{¬}{\neg}
\newunicodechar{∨}{\vee}
\newunicodechar{∧}{\wedge}
\newunicodechar{∪}{\cup}
\newunicodechar{∩}{\cap}
\newunicodechar{⋁}{\bigvee}
\newunicodechar{⋀}{\bigwedge}
\newunicodechar{⋃}{\bigcup}
\newunicodechar{⋂}{\bigcap}
\newunicodechar{⟂}{\perp}
\newunicodechar{∥}{\parallel}
\newunicodechar{∦}{\nparallel}
\newunicodechar{𝚤}{\imath}
\newunicodechar{𝚥}{\jmath}
\newunicodechar{⇔}{\Leftrightarrow}
\newunicodechar{⇕}{\Updownarrow}
\newunicodechar{⇐}{\Leftarrow}
\newunicodechar{⇒}{\Rightarrow}
\newunicodechar{⇑}{\Uparrow}
\newunicodechar{⇓}{\Downarrow}
\newunicodechar{↔}{\leftrightarrow}
\newunicodechar{↕}{\updownarrow}
\newunicodechar{←}{\leftarrow}
\newunicodechar{→}{\rightarrow}
\newunicodechar{↑}{\uparrow}
\newunicodechar{↓}{\downarrow}
\newunicodechar{⟷}{\longleftrightarrow}
\newunicodechar{⟵}{\longleftarrow}
\newunicodechar{⟶}{\longrightarrow}
\newunicodechar{⇇}{\leftleftarrows}
\newunicodechar{⇉}{\rightrightarrows}
\newunicodechar{⇈}{\upuparrows}
\newunicodechar{⇊}{\downdownarrows}
\newunicodechar{⟺}{\Longleftrightarrow}
\newunicodechar{⟸}{\Longleftarrow}
\newunicodechar{⟹}{\Longrightarrow}
\newunicodechar{↦}{\mapsto}
\newunicodechar{↤}{\mapsfrom}
\newunicodechar{⟼}{\longmapsto}
\newunicodechar{⟻}{\longmapsfrom}
\newunicodechar{⟾}{\Longmapsto}
\newunicodechar{⟽}{\Longmapsfrom}
\newunicodechar{↗}{\nearrow}
\newunicodechar{↖}{\nwarrow}
\newunicodechar{↘}{\searrow}
\newunicodechar{↙}{\swarrow}
\newunicodechar{↩}{\hookleftarrow}
\newunicodechar{↪}{\hookrightarrow}
\newunicodechar{↶}{\curvearrowleft}
\newunicodechar{↷}{\curvearrowright}
\newunicodechar{↺}{\circlearrowleft}
\newunicodechar{↻}{\circlearrowright}
\newunicodechar{↫}{\looparrowleft}
\newunicodechar{↬}{\looparrowright}
\newunicodechar{⇋}{\leftrightharpoons}
\newunicodechar{⇌}{\rightleftharpoons}
\newunicodechar{↼}{\leftharpoonup}
\newunicodechar{↽}{\leftharpoondown}
\newunicodechar{⇀}{\rightharpoonup}
\newunicodechar{⇁}{\rightharpoondown}
\newunicodechar{↿}{\upharpoonleft}
\newunicodechar{↾}{\upharpoonright}
\newunicodechar{⇃}{\downharpoonleft}
\newunicodechar{⇂}{\downharpoonright}
\newunicodechar{𝔸}{\mathbb{A}}
\newunicodechar{𝔹}{\mathbb{B}}
\newunicodechar{ℂ}{\mathbb{C}}
\newunicodechar{𝔻}{\mathbb{D}}
\newunicodechar{𝔼}{\mathbb{E}}
\newunicodechar{𝔽}{\mathbb{F}}
\newunicodechar{𝔾}{\mathbb{G}}
\newunicodechar{ℍ}{\mathbb{H}}
\newunicodechar{𝕀}{\mathbb{I}}
\newunicodechar{𝕁}{\mathbb{J}}
\newunicodechar{𝕂}{\mathbb{K}}
\newunicodechar{𝕃}{\mathbb{L}}
\newunicodechar{𝕄}{\mathbb{M}}
\newunicodechar{ℕ}{\mathbb{N}}
\newunicodechar{𝕆}{\mathbb{O}}
\newunicodechar{ℙ}{\mathbb{P}}
\newunicodechar{ℚ}{\mathbb{Q}}
\newunicodechar{ℝ}{\mathbb{R}}
\newunicodechar{𝕊}{\mathbb{S}}
\newunicodechar{𝕋}{\mathbb{T}}
\newunicodechar{𝕌}{\mathbb{U}}
\newunicodechar{𝕍}{\mathbb{V}}
\newunicodechar{𝕎}{\mathbb{W}}
\newunicodechar{𝕏}{\mathbb{X}}
\newunicodechar{𝕐}{\mathbb{Y}}
\newunicodechar{ℤ}{\mathbb{Z}}
\newunicodechar{𝒜}{\mathcal{A}}
\newunicodechar{ℬ}{\mathcal{B}}
\newunicodechar{𝒞}{\mathcal{C}}
\newunicodechar{𝒟}{\mathcal{D}}
\newunicodechar{ℰ}{\mathcal{E}}
\newunicodechar{ℱ}{\mathcal{F}}
\newunicodechar{𝒢}{\mathcal{G}}
\newunicodechar{ℋ}{\mathcal{H}}
\newunicodechar{ℐ}{\mathcal{I}}
\newunicodechar{𝒥}{\mathcal{J}}
\newunicodechar{𝒦}{\mathcal{K}}
\newunicodechar{ℒ}{\mathcal{L}}
\newunicodechar{ℳ}{\mathcal{M}}
\newunicodechar{𝒩}{\mathcal{N}}
\newunicodechar{𝒪}{\mathcal{O}}
\newunicodechar{𝒫}{\mathcal{P}}
\newunicodechar{𝒬}{\mathcal{Q}}
\newunicodechar{ℛ}{\mathcal{R}}
\newunicodechar{𝒮}{\mathcal{S}}
\newunicodechar{𝒯}{\mathcal{T}}
\newunicodechar{𝒰}{\mathcal{U}}
\newunicodechar{𝒱}{\mathcal{V}}
\newunicodechar{𝒲}{\mathcal{W}}
\newunicodechar{𝒳}{\mathcal{X}}
\newunicodechar{𝒴}{\mathcal{Y}}
\newunicodechar{𝒵}{\mathcal{Z}}
\newunicodechar{𝕬}{\mathfrak{A}}
\newunicodechar{𝕭}{\mathfrak{B}}
\newunicodechar{𝕮}{\mathfrak{C}}
\newunicodechar{𝕯}{\mathfrak{D}}
\newunicodechar{𝕰}{\mathfrak{E}}
\newunicodechar{𝕱}{\mathfrak{F}}
\newunicodechar{𝕲}{\mathfrak{G}}
\newunicodechar{𝕳}{\mathfrak{H}}
\newunicodechar{𝕴}{\mathfrak{I}}
\newunicodechar{𝕵}{\mathfrak{J}}
\newunicodechar{𝕶}{\mathfrak{K}}
\newunicodechar{𝕷}{\mathfrak{L}}
\newunicodechar{𝕸}{\mathfrak{M}}
\newunicodechar{𝕹}{\mathfrak{N}}
\newunicodechar{𝕺}{\mathfrak{O}}
\newunicodechar{𝕻}{\mathfrak{P}}
\newunicodechar{𝕼}{\mathfrak{Q}}
\newunicodechar{𝕽}{\mathfrak{R}}
\newunicodechar{𝕾}{\mathfrak{S}}
\newunicodechar{𝕿}{\mathfrak{T}}
\newunicodechar{𝖀}{\mathfrak{U}}
\newunicodechar{𝖁}{\mathfrak{V}}
\newunicodechar{𝖂}{\mathfrak{W}}
\newunicodechar{𝖃}{\mathfrak{X}}
\newunicodechar{𝖄}{\mathfrak{Y}}
\newunicodechar{𝖅}{\mathfrak{Z}}

\DeactivateWarningFilters[pdflatex-unicode-math]

\usepackage{siunitx}
\usepackage{fnbreak}
\usepackage{booktabs}
\usepackage{gincltex}
\usepackage{grffile}
\usepackage{import}
\usepackage{tikz}
\usepackage{acro}
\usepackage[useregional, en-GB]{datetime2}
\usepackage{csquotes}
\usepackage{cleveref}

\graphicspath{{./res/}}

\crefname{figure}{Fig.}{Figs.}
\crefname{table}{Table}{Tables}
\renewcommand*{\vec}{\mathbfit}
\newcommand*{\imag}{\mathrm{i}}
\newcommand*{\euler}{\mathrm{e}}
\newcommand*{\Ms}{\ensuremath{\mathrm{M}_\odot}}
\DeclareSIUnit{\Ms}{\text{\ensuremath{\mathrm{M}_\odot}}}
\newcommand*{\textcode}[1]{\textsc{\small #1}}

\newcommand*{\tmp}{}
\newcommand*{\q}{\enquote}
\newcommand*{\email}[1]{\href{mailto:#1}{\texttt{#1}}}

\crefname{algorithm}{step}{steps}%
\creflabelformat{algorithm}{(#2#1#3)}%

\DeclareSIUnit{\pc}{pc}
\DeclareSIUnit{\kpc}{\kilo\pc}
\DeclareSIUnit{\Mpc}{\mega\pc}
\DeclareSIUnit{\c}{\text{\ensuremath{c}}}
\DeclareSIUnit{\hHubble}{\text{\ensuremath{h}}}
\DeclareSIUnit{\CPU}{CPU}

\newcommand*{\ml}{\mleft}
\newcommand*{\mr}{\mright}
\newcommand*{\mtext}[1]{{\operatorfont #1}}
\newcommand*{\phys}{{\mtext{phys}}}
\let\Re=\relax
\DeclareMathOperator{\Re}{Re}

\DeclareMathOperator{\FFT}{FFT}

\def\diffd{\mathrm{d}}
\DeclareDocumentCommand\differential{ o g d() }{ 
	\IfNoValueTF{#2}{
		\IfNoValueTF{#3}
		{\diffd\IfNoValueTF{#1}{}{^{#1}}}
		{\mathinner{\diffd\IfNoValueTF{#1}{}{^{#1}}\argopen(#3\argclose)}}
	}
	{\mathinner{\diffd\IfNoValueTF{#1}{}{^{#1}}#2} \IfNoValueTF{#3}{}{(#3)}}
}
\DeclareDocumentCommand\dd{}{\differential} 

\newcommand*{\AxiREPO}{\textcode{AxiREPO}}
\newcommand*{\AREPO}{\textcode{AREPO}}
\newcommand*{\DisPerSE}{\textcode{DisPerSE}}
\newcommand*{\NGenIC}{\textcode{N-GenIC}}
\newcommand*{\axionCAMB}{\textcode{axionCAMB}}

\DeclareAcronym{AMR}{short=AMR, long=adaptive mesh refinement}
\DeclareAcronym{ALP}{short=ALP, long=axion-like particle}
\DeclareAcronym{CDM}{short=CDM, long=cold dark matter}
\DeclareAcronym{CiC}{short=CiC, long=cloud-in-cell}
\DeclareAcronym{FDM}{short=FDM, long=fuzzy dark matter}
\DeclareAcronym{FFT}{short=FFT, long=Fast Fourier Transform}
\DeclareAcronym{FFTW}{short=FFTW, long=Fastest Fourier Transform in the West}
\DeclareAcronym{FoF}{short=FoF, long=friends-of-friends}
\DeclareAcronym{FRW}{short=FRW, long=Friedmann–Robertson–Walker}
\DeclareAcronym{HMF}{short=HMF, long=halo mass function}
\DeclareAcronym{HPC}{short=HPC, long=high-performance computing}
\DeclareAcronym{IC}{
	short=IC, long=initial condition,
	foreign-plural={}
}
\DeclareAcronym{LCDM}{short=$Λ$CDM, long=cold dark matter and a cosmological constant}
\DeclareAcronym{MPCDF}{short=MPCDF, long=Max Planck Computing and Data Facility}
\DeclareAcronym{MPI}{short=MPI, long=Message Passing Interface}
\DeclareAcronym{NFW}{short=NFW, long=Navarro–Frenk–White}
\DeclareAcronym{QCD}{short=QCD, long=quantum chromodynamics}
\DeclareAcronym{QFT}{short=QFT, long=quantum field theory}
\DeclareAcronym{SIDM}{short=SIDM, long=self-interacting dark matter}
\DeclareAcronym{SP}{short=SP, long=Schrödinger–Poisson}
\DeclareAcronym{SPH}{short=SPH, long=smoothed-particle hydrodynamics}
\DeclareAcronym{UFD}{short=UFD, long=ultra-faint dwarf}
\DeclareAcronym{WDM}{short=WDM, long=warm dark matter}
\DeclareAcronym{WIMP}{
	short=WIMP, long=weakly interacting massive particle,
	foreign-plural={}
}

\title[The halo mass function and filaments in fuzzy dark matter models]{%
	The halo mass function and filaments in full cosmological simulations with fuzzy dark matter%
}

\author[S.~May and V.~Springel]{%
	Simon May$^{1}$\thanks{E-mail: \email{simon.may@mpa-garching.mpg.de}} and
	Volker Springel$^{1}$
	\\
	$^{1}$Max-Planck-Institut für Astrophysik, Karl-Schwarzschild-Straße 1, 85741 Garching, Germany
}

\date{Accepted XXX. Received YYY; in original form ZZZ}

\pubyear{2022}

\begin{document}

\label{firstpage}
\pagerange{\pageref{firstpage}--\pageref{lastpage}}
\maketitle

\begin{abstract}
	Fuzzy dark matter (FDM) is a dark matter candidate consisting of ultra-light scalar particles with masses around $10^{-22}\,\mathrm{eV}/c^2$, a regime where cold bosonic matter behaves as a collective wave rather than individual particles.
	It has increasingly attracted attention due to its rich phenomenology on astrophysical scales, with implications for the small-scale tensions present within the standard cosmological model, $\Lambda$CDM.
	Although constraints on FDM are accumulating in many different contexts, very few have been verified by self-consistent numerical simulations.
	We present new large numerical simulations of cosmic structure formation with FDM, solving the full Schrödinger–Poisson (SP) equations using the {\AxiREPO} code, which implements a pseudo-spectral numerical method.
	Combined with our previous simulations, they allow us to draw a four-way comparison of matter clustering, contrasting results (such as power spectra) for each combination of initial conditions (FDM vs.\ CDM) and dynamics (Schrödinger–Poisson vs.\ $N$-body).
	By disentangling the impact of initial conditions and non-linear dynamics in this manner, we can gauge the validity of approximate methods used in previous works, such as ordinary $N$-body simulations with an FDM initial power spectrum.
	Due to the comparatively large volume achieved in our FDM simulations, we are able to measure the FDM halo mass function from full wave simulations for the first time, and compare to previous results obtained using analytic or approximate approaches.
	We find that, due to the cut-off of small-scale power in the FDM power spectrum, haloes are linked via continuous, smooth, and dense filaments throughout the entire simulation volume (unlike for the standard $\Lambda$CDM power spectrum), posing significant challenges for reliably identifying haloes.
	We also investigate the density profiles of these filaments and compare to their $\Lambda$CDM counterparts.
\end{abstract}



\section{Introduction}
\label{sec:introduction}

A number of tensions between theory and observations on small cosmological scales, such as the \q{cusp–core problem}, the \q{missing satellite problem}, or the \q{too-big-to-fail problem} \citep{2015PNAS..11212249W, 2017Galax...5...17D, 2011MNRAS.415L..40B} have sometimes raised questions about the validity of the \q{standard cosmological model} based on \ac{LCDM}, which has otherwise been extremely successful in describing a wide variety of cosmological observations across a broad range of physical scales\citep[e.\,g.][]{2012AnP...524..507F, 2016PDU....12...56B}. Distinguishing effects due to baryonic physics from genuine failures of \ac{LCDM} has proven to be a significant challenge, however \citep[e.\,g.][]{Santos-Santos2020, Grand2021, Sales2022}.  Further, \ac{CDM} as a model only makes predictions in the context of cosmology, with countless possible implementations of galaxy formation physics \citep[see][for a recent rewiew]{Vogelsberger2020}.

Although large-scale structure studies have solidified the correspondence of observations with the behaviour of \ac{CDM}, dark matter models can still deviate on less well constrained, smaller (i.\,e.\ galactic) scales.  In this regard, \ac{FDM} offers a wide range of new phenomena which have an impact on some of the \q{small-scale problems} \citep[see e.\,g.][for reviews]{2016PhR...643....1M, 2017PhRvD..95d3541H, 2020PrPNP.11303787N, 2021ARAA..59..247H, 2021AARv..29....7F}.  Due to the small particle masses, interesting wave effects occur which are unique to this class of dark matter models.  Early numerical simulations have already shown that ultra-light scalars form cores in the centres of dark matter haloes \citep{2014NatPh..10..496S}, possibly explaining observed dwarf galaxy rotation curves \citep[but see][]{2020ApJ...904..161B}.  In addition, the cut-off in the \ac{FDM} transfer function, suppressing small-scale power in a similar fashion as \ac{WDM}, has the potential to solve issues like the \q{missing satellite problem}.
In a sense, \ac{FDM} combines properties similar to the features of \ac{WDM} (suppression on small scales) and \ac{SIDM} (cored density profiles) with respect to the small-scale challenges mentioned above \citep{2019PrPNP.104....1B, 2018PhR...730....1T}.

However, \ac{FDM} also exhibits a number of additional unique phenomena.
Objects other than cores, such as quantized vortices, are another differentiating feature with interesting prospects for detection \citep{2021JCAP...01..011H}, and relative fluctuations of order one in the density field at the scale of the \ac{FDM} wavelength can have a strong impact on visible matter e.\,g.\ through dynamical heating.
Furthermore, light (pseudo-)scalar particles are a common feature of theories in particle physics, from the original axion in \ac{QCD} to a plethora of axion-like particles predicted by unified and early-universe theories such as string theories \citep{2016PhR...643....1M}.

After \ac{FDM} garnered great interest due to motivation from astrophysics and particle physics, as summarised above, constraints on the mass $m$ of the scalar particles at values around $mc^2 ≳ \SI{e-21}{\eV}$ accumulated using many different contexts and observables for some time \citep{2021AARv..29....7F}.
It has turned out that the scaling relations of \ac{FDM} halo cores are difficult to reconcile with observations when assuming a single value for the particle mass $m$ \citep{2020ApJ...904..161B}, and although this does not rule out \ac{FDM} as a dark matter candidate by itself, it does weaken its motivation as a possible solution to the \q{cusp–core problem}.
Relatedly, recent results from dynamical modelling of \ac{UFD} galaxies (where \ac{FDM} cores become increasingly large) has been used to claim rather strong constraints on $m$, with values up to $≈ \SI{e-19}{\eV}$ \citep{2021ApJ...912L...3H, 2021AA...651A..80Z}.
However, these analyses do not take the significant scatter in the \ac{FDM} core–halo mass relation, discovered in cosmological simulations, into account \citep{2022MNRAS.511..943C}, which would weaken these bounds.
A very recent result from \citet{2022arXiv220305750D}, claiming $mc^2 > \SI{3e-19}{\eV}$ due to the heating effect of stellar orbits caused by potential fluctuations in \ac{FDM} haloes, has more serious implications, since it does not depend on the uncertain core–halo relation.

However, most of these constraints (including the most recent) have in common that they have not yet been verified by self-consistent numerical simulations, instead relying on approximate, idealised, or simplified numerical and analytic approaches.
While cosmological \ac{FDM} simulations have been carried out using a variety of numerical methods, many attempts were quite limited in scope
\citep[][table~1]{2018FrASS...5...48Z, 2021MNRAS.504.2391L}, so that the effects of \ac{FDM} in (mildly) non-linear regimes of structure formation are still poorly understood compared to \ac{CDM}.

Apart from an overall still lower level of research attention, an important reason impeding insight into \ac{FDM} lies in the very large computational costs incurred when numerically solving the corresponding equations of motion – these costs are much higher than the ones associated with corresponding \ac{LCDM} calculations.
Due to the computational requirements, the cosmological volumes studied in simulations with the full \ac{FDM} \ac{SP} dynamics have been especially limited \citep{2009ApJ...697..850W, 2014NatPh..10..496S, 2018PhRvD..98d3509V, 2020MNRAS.494.2027M}.
Although recent advances in hybrid numerical techniques have made it feasible to embed simulated \ac{FDM} haloes within much larger simulated boxes \citep{2022PhRvL.128r1301S}, the fundamental issues driving the cost in the regions where the full \ac{FDM} equations of motion are treated remain.
In this work, we carry out simulations that smoothly connect the non-linear state reached in isolated \ac{FDM} haloes to the still linear large-scale structure, thereby bridging, in particular, the regime of mildly non-linear evolution where differences in the temporal evolution compared to \ac{CDM} can be expected.
We complement our previous simulations \citep{2021MNRAS.506.2603M} by including the self-consistent transfer function expected for an \ac{FDM} cosmology.
As before, we carry out very large \ac{FDM} simulations with a spectral method on a uniform grid, which fully retains the quantum-mechanical effects.
We compare and contrast with our previous results for central measures of matter clustering, namely the power spectrum and the \ac{HMF}, and highlight the unique challenges and phenomena related to filaments which arise when using an \ac{FDM} initial power spectrum.

While methods which forego a treatment of the full wave dynamics have been able to conduct simulations with volumes much closer to those attainable using traditional $N$-body and \ac{SPH} approaches for \ac{CDM} \citep{2016ApJ...818...89S, 2016PhRvD..94l3523V, 2018ApJ...863...73Z, 2018MNRAS.478.3935N, 2019MNRAS.482.3227N, 2021MNRAS.501.1539N}, these do not capture inherent wave phenomena such as interference effects, which can have a significant impact on the overall evolution at least on small scales \citep{2019PhRvD..99f3509L}, leaving the validity of results obtained in this way unclear in the absence of similar computations solving the fundamental wave equations.
While the hybrid method from \citet{2022PhRvL.128r1301S} improves upon the computational limitations, by its nature it also does not incorporate the full \ac{FDM} evolution, and can only reproduce it in a statistical sense.
In particular, while all simulations can easily incorporate the impact of the suppressed small-scale power spectrum present with \ac{FDM} in the initial conditions, such methods either lack the wave nature of \ac{FDM} entirely or only approximate it.
Using our simulations, we are now equipped to fully clarify the reliability of such approximative results, disentangling the two essential physical differences distinguishing \ac{FDM} from \ac{CDM} in cosmological numerical simulations: the dynamics (equations of motion) and the \acp{IC}.

The paper is structured as follows. In \cref{sec:theory}, we concisely summarize the theoretical background of \ac{FDM} cosmologies, while in \cref{sec:numerics} we detail our numerical methodology. In \cref{sec:power-spectrum} we compare matter clustering in \ac{FDM} and \ac{CDM} cosmologies at the level of the power spectrum, for different sets of initial conditions. We then turn in \cref{sec:halo-mass-function} to a discussion of the challenges involved in measuring the halo mass function in self-consistent cosmological \ac{FDM} simulation. In \cref{sec:filaments} we consider the density structure of filaments in \ac{FDM} as compared to \ac{CDM}, while we return to halo profiles in \cref{sec:halo-profiles}. Finally, we present a summary of our findings and our conclusions in \cref{sec:conclusions}.

\section{Theoretical background}
\label{sec:theory}

\Acl{FDM} is a particular form of scalar field dark matter with an \q{unusual} choice of the scalar field particle mass $m$.
It is described by the simple scalar field action
\renewcommand*{\tmp}{\ensuremath{%
	- \frac{1}{2} \frac{m^2 c^2}{ℏ^2} ϕ(x)^2 - \frac{λ}{ℏ^2 c^2} ϕ(x)^4%
}}
\begin{multline}
	\label{eq:action}
	S =
	\frac{1}{ℏc^2} ∫\! \dd[4]{x} √{-g} \, \ml(\frac{1}{2} g^{μν} \ml(∂_μ ϕ(x)\mr) \ml(∂_ν ϕ(x)\mr) \vphantom{\tmp}\mr.
	\\
	\ml. \tmp \mr),
\end{multline}
with the metric $g^{μν}$ and its determinant $g$, a real scalar field $ϕ$, its mass $m$, and a self-interaction coupling strength $λ$.\footnote{%
	$c$ and $ℏ$ are the speed of light in vacuum and the reduced Planck constant, and are explicitly included in all equations.%
}
In this most fundamental description of quantum field theory in a curved spacetime, $ϕ$ is a quantum field.
This action is to be understood in the context of quantum field theory in a curved spacetime, i.\,e.\ $ϕ$ is a (\q{second-quantised}) quantum field, although for the purposes of numerical calculations, we will only consider the \q{classical} and non-relativistic limits.
As in \citet{2021MNRAS.506.2603M}, we will not consider self-interactions, i.\,e.\ $λ = 0$.

As mentioned above, the distinguishing feature of the \ac{FDM} model is the value of the mass $m$, which is around $m c^2 ≈ \SI{e-22}{\eV}$, making this an \emph{ultra-light} scalar field.
This is in stark contrast to most realizations of \ac{CDM}, such as \ac{WIMP} models (which include scalar field dark matter models), which feature particle masses in the range of \SI{100}{\GeV}–\si{\TeV}.
A consequence of this wildly different mass regime is that the \ac{FDM}  scalar field is better described as waves instead of individual particles, making \ac{FDM} an example of a \emph{wave dark matter} model, with associated striking differences in phenomenology.
The reason is that, for such ultra-light values of the particle mass, the resulting particle number densities are extremely large, such that the quantum-mechanical de Broglie wavelength will be much larger than the inter-particle separation.
Correspondingly, a collective wave description becomes much more appropriate than dealing with individual particles.
Notice that such dark matter models require a non-thermal production mechanism to remain \q{cold}, in which case bosons (like the scalar particles of \ac{FDM}) can form a Bose–Einstein condensate.

As mentioned above, we will assume non-relativistic approximations appropriate for simulations of cosmic structure formation, where all velocities are $≪ c$, and the considered scales are smaller than the Hubble horizon $c / H_0$.
Rewriting the real scalar field $ϕ$ in terms of a complex variable $ψ$,
\begin{equation}
	ϕ
	= \frac{1}{2} √{\frac{ℏ^3 c}{2m}} \Re\ml(ψ \euler^{-\imag\frac{mc^2}{ℏ} t}\mr)
	= √{\frac{ℏ^3 c}{2m}} \ml(ψ \euler^{-\imag\frac{mc^2}{ℏ} t} + ψ^* \euler^{\imag\frac{mc^2}{ℏ} t}\mr)
	,
\end{equation}
and taking the non-relativistic limit in the Newtonian gauge yields the equations of motions for \acl{FDM}, the \ac{SP} equations:\footnote{%
	Since \cref{eq:fdm-schroedinger}, despite being formally identical to the Schrödinger equation from single-particle quantum mechanics, has a very different physical meaning and is in fact a Gross–Pitaevskii equation \citep{Gross1961, Pitaevskii1961}, \cref{eq:fdm-schroedinger,eq:fdm-poisson} are technically the Gross–Pitaevskii–Poisson system of equations.
	However, we will use the more common term \q{\acl{SP} equations} here.%
}
\begin{align}
	\label{eq:fdm-schroedinger}
	\imag ℏ ∂_t ψ(t, \vec{x})
	&= -\frac{ℏ^2}{2ma(t)^2} ∇^2 ψ(t, \vec{x}) + \frac{m}{a(t)} Φ ψ(t, \vec{x}),
	\\
	\label{eq:fdm-poisson}
	∇^2 Φ(t, \vec{x})
	&= 4πGm \ml(|ψ(t, \vec{x})|^2 - ⟨|ψ|^2⟩(t)\mr)
	,
\end{align}
where $a$ is the cosmological scale factor, $Φ$ is the Newtonian gravitational potential, $G$ is Newton's constant, and the angle brackets in $⟨|ψ|^2⟩$ indicate a spatial average.
These are the equations of motions whose non-linear time evolution is solved in our numerical simulations.
In the above, all quantities and coordinates are given in \q{comoving} form, to factor out the dependence on the scale factor $a(t)$ as much as possible.\footnote{%
	All quantities will be given in terms of comoving lengths unless specified otherwise.%
}
Explicitly, comoving and \q{physical} quantities are related as follows:
\begin{equation}
	\vec{x} = a^{-1} \vec{x}_{\phys},
	\quad
	∇ = a∇_{\phys},
	\quad
	ψ = a^{\sfrac{3}{2}} ψ_{\phys},
	\quad
	Φ = aΦ_{\phys}.
\end{equation}

As we have assumed a production mechanism that can be described as a Bose–Einstein condensate, with almost all of the scalar particles in their quantum-mechanical single-particle ground states and large number densities, we can make use of the mean field approximation, also called the \q{classical} limit or wave limit, where the particles behave collectively and coherently.
The \q{wave function} $ψ$ in \cref{eq:fdm-schroedinger}, which is now simply a complex number-valued function (instead of a quantum field), is then understood as the single macroscopic amplitude of the \ac{FDM} waves, with the mass density given by
\begin{equation}
	\label{eq:mass-density}
	ρ = m|ψ|^2
	.
\end{equation}
This is why \ac{FDM} is often called a \q{classical theory}.\footnote{%
	The distinction is irrelevant for the physical and phenomenological consequences of \ac{FDM}, but often seems to spark debate.
	We would like to note that, although \ac{FDM} may be described by what is formally a classical field theory, the underlying \emph{physical} phenomena, such as matter waves and Bose–Einstein condensation, were not known in classical physics.%
}

As in quantum mechanics, or generally for diffusion equations, \cref{eq:fdm-schroedinger} obeys the continuity equation
\begin{gather}
	\label{eq:continuity}
	∂_t ρ + ∇ ⋅ ρ \vec{v} = 0,
	\intertext{with the momentum density}
	\label{eq:momentum-density}
	ρ \vec{v}
	= \frac{ℏ}{2\imag} \ml(ψ^* ∇ ψ - ψ ∇ ψ^*\mr).
\end{gather}
The momentum density $ρ \vec{v}$ can also be used to define the bulk peculiar velocity field $\vec{v}$, although this is problematic in the case of $ρ = 0$.
Separating the complex wave amplitude $ψ$ into its absolute value $√{ρ / m}$ and complex phase $θ$,
\begin{equation}
	\label{eq:wave-function-polar}
	ψ = √{\frac{ρ}{m}} \euler^{\imag θ}
	,
\end{equation}
yields a simpler form of \cref{eq:momentum-density}:
\begin{equation}
	\label{eq:velocity}
	\vec{v} = \frac{ℏ}{m} ∇θ
	,
\end{equation}
i.\,e.\ the velocity is given by the gradient of the wave amplitude's phase.
Again, however, this is ill-defined when $ψ = ρ = 0$, in which case the phase $θ$ is undefined.
It should be noted that due to the presence of wave interference, the case $ψ = 0$ is actually quite common (occurring wherever destructive interference takes place) and cannot be neglected as perhaps for non wave-like dark matter models.

In addition to the mass (as implied by the continuity equation, \cref{eq:continuity}), the total energy, given by
\begin{align}
	\label{eq:energy}
	E
	&= ∫ \dd[3]{x} \ml(\frac{ℏ^2}{2m^2} |∇ψ|^2 + \frac{1}{2} Φ |ψ|^2\mr)
	\notag
	\\
	&= ∫ \dd[3]{x} \frac{ℏ^2}{2m^2} (∇√ρ)^2
	+ ∫ \dd[3]{x} \frac{1}{2} ρv^2
	+ ∫ \dd[3]{x} \frac{1}{2} ρΦ
	\notag
	\\
	&= T_ρ + T_v + V
\end{align}
is conserved \citep{2017MNRAS.471.4559M}.
The kinetic energy $T = T_ρ + T_v$ is made up of the \q{bulk} kinetic energy $T_v$, and a gradient energy term $T_ρ$, while $V$ is the standard gravitational potential energy.

The separation of the wave amplitude into two real variables and their connection with the density \eqref{eq:mass-density} and velocity \eqref{eq:velocity} allows a hydrodynamical interpretation analogous to that of \citet{Madelung1927} in quantum mechanics, although its validity is problematic for destructive interference.
\Ac{FDM}'s resistance against gravitational collapse on small scales, which can be interpreted as an analogue to the Heisenberg uncertainty principle in the \ac{SP} formulation, then manifests itself as an explicit additional \q{quantum pressure} appearing in the analogue of the hydrodynamical Euler equation.

There are two important length scales which serve as indicators of \ac{FDM} wave phenomena, both of which are determined by the constant $ℏ / m$, which is the only independent parameter in the equations of motion, \cref{eq:fdm-schroedinger,eq:fdm-poisson}.
The first is the de Broglie wavelength
\begin{equation}
	\label{eq:de-broglie}
	λ_{\mtext{dB}} = \frac{2πℏ}{mv}
	,
\end{equation}
which has the same meaning as in quantum mechanics and indicates the length scale on which density fluctuations of order one occur.
The second is the \ac{FDM} analogue of the Jeans length.
In linear perturbation theory, this is the length scale where the gravitational attraction balances the \q{quantum pressure} resisting collapse, such that perturbations on larger scales grow, while smaller ones oscillate.
Expressed as a wave number $k_{\mtext{J}}$, this is \citep{2000PhRvL..85.1158H}
\begin{equation}
	\label{eq:jeans}
	k_{\mtext{J}}
	= \ml(\frac{6\, Ω_{\mtext{m}}}{1 + z}\mr)^{\sfrac{1}{4}} \ml(\frac{m H_0}{ℏ}\mr)^{\sfrac{1}{2}}
\end{equation}
as a function of the redshift $z$, with the Hubble parameter $H_0$, and the cosmic matter density parameter $Ω_{\mtext{m}}$.

\section{Numerical methodology}
\label{sec:numerics}

Our numerical simulations are performed using the same setup as in \citet{2021MNRAS.506.2603M}.
We present two types of simulations:
\begin{inparaenum}[(1)]
	\item \Ac{FDM} simulations using the {\AxiREPO} code\footnote{%
		The {\AxiREPO} code will be made public in the near future (including the halo finder).%
	} first introduced in \citet{2021MNRAS.506.2603M}, which is implemented as a module in {\AREPO} \citep{2010MNRAS.401..791S} and numerically solves the \ac{SP} equations, \cref{eq:fdm-schroedinger,eq:fdm-poisson}, using a pseudo-spectral method, and
	\item \q{standard} $N$-body \ac{CDM} simulations using the unmodified {\AREPO} code, which implements the TreePM numerical method to solve the Vlasov–Poisson equations.
\end{inparaenum}
For each type of simulation, we additionally compare and contrast two different kinds of \acp{IC}, corresponding to an \ac{FDM} or \ac{CDM} universe (see \cref{sec:ics}).

As in our previous work, the simulation volume consists of a cubic box of side length $L$ with periodic boundary conditions, sampling the matter distribution in the universe.
The box is filled with dark matter (\ac{FDM} or \ac{CDM}) whose average comoving density is the cosmic mean background matter density
\begin{equation}
	⟨ρ⟩
	= ρ_{\mtext{m}}
	= Ω_{\mtext{m}} ρ_{\mtext{crit}}
	= Ω_{\mtext{m}} \frac{3 H_0^2}{8πG}
	.
\end{equation}
In the \ac{FDM} case (pseudo-spectral \ac{SP} solver), the fields $ψ$ (\q{wave function}) and $Φ$ (\q{potential}) are discretised on a uniform Cartesian grid with $N^3$ points, enabling the use of the \ac{FFT}.
In the \ac{CDM} case, as usual, the phase space of dark matter particles is sampled using (much) more massive \q{simulation particles}, whose trajectories are evolved using Newtonian gravitational dynamics.

Concerning \ac{FDM} \ac{SP} simulations, it is important to make note of the tremendous computational requirements involved.
Firstly, for the pseudo-spectral method, ensuring the absence of \q{aliasing} in the complex exponentials imposes a restriction on the size of the time step $Δt$,
\begin{equation}
	\label{eq:time-step}
	Δt < \min\ml(
		\frac{4}{3π} \frac{m}{ℏ} a^2 Δx^2,\;
		2π \frac{ℏ}{m} a \frac{1}{|Φ_{\mtext{max}}|}
	\mr)
	,
\end{equation}
where $Δx = L / N$ is the spatial grid resolution and $Φ_{\mtext{max}}$ is the maximum value of the potential.
Although the exact form of \cref{eq:time-step} is specific to the pseudo-spectral method, the dependence $Δt ∝ Δx^2$ seems to hold for all numerical approaches to the \ac{SP} system of equations,
\cref{eq:fdm-schroedinger,eq:fdm-poisson}, and serves as an illustration of the Schrödinger equation's nature as a diffusion equation.

Secondly, a constraint on the validity of the spatial discretisation emerges from the relationship between the velocity field and the gradient of the wave function's phase (\cref{eq:velocity}), which implies that velocities $v$ larger than a certain value $v_{\mtext{max}}$ cannot be represented in a simulation with spatial resolution $Δx$:
\begin{equation}
	\label{eq:velocity-criterion-v}
	v_{\mtext{max}} = \frac{ℏ}{m} \frac{π}{Δx}
	.
\end{equation}
In other words, the \q{worst} spatial resolution $Δx_{\mtext{max}}$ that still yields acceptable results is (roughly) given by the de Broglie wave length corresponding to $v_{\mtext{max}}$:
\begin{equation}
	\label{eq:velocity-criterion-x}
	\begin{aligned}
		Δx_{\mtext{max}} &=
		\frac{πℏ}{mv_{\mtext{max}}}
		= \frac{1}{2} λ_{\mtext{dB}}(v_{\mtext{max}})
		\\
		⇔
		Δx &< \frac{πℏ}{mv_{\mtext{max}}}
		.
	\end{aligned}
\end{equation}
Or, rephrasing yet again, the de Broglie wavelength must be resolved for all velocities appearing in the simulation.
As with the time step criterion, similar considerations mandate resolving the de Broglie wavelength also for other numerical methods for the \ac{SP} equations, yielding a similar constraint.

Combined, the time step criterion \eqref{eq:time-step}, $Δt ∝ Δx^2$, and the velocity criterion \eqref{eq:velocity-criterion-x}, $Δx ∝ 1 / v$, render simulations of large cosmological objects or representative cosmological volumes prohibitively expensive.
Contrary to $N$-body simulations, the additional requirement on the spatial resolution makes it impossible to resort to coarser, low-resolution simulations while still obtaining valid results on all (large) scales that are still resolved.
Instead, a lack of sufficient resolution will affect results even on the largest scales and within the linear regime of structure formation, as shown in \citet{2021MNRAS.506.2603M}.

Although numerical methods have been developed for \ac{FDM} which are in principle more versatile or sophisticated, such as \ac{AMR} or hybrid techniques \citep{2022PhRvL.128r1301S}, the unique computational demands of \ac{FDM} have the effect that, except for specific cases where very high resolution is desired in a small region (\q{zoom-in}), these methods still suffer from the tremendous cost involved in \ac{FDM} simulations.
While it may be argued that a uniform grid is wasteful in regions where a lower resolution would be sufficient, the velocity criterion \eqref{eq:velocity-criterion-x} actually imposes significant demands on resolution even in low-density regions.
The pseudo-spectral method used here, although not very flexible due to its limitation to a single uniform resolution, nevertheless has the advantage of unmatched accuracy in the spatial integration (at the given level of resolution), in addition to its use of a very simple and optimized algorithm in the form of the \ac{FFT}, so it is at present one of the best ways to address the unique challenges of the \ac{SP} system of equations.

\subsection{Initial conditions}
\label{sec:ics}

As in \citet{2021MNRAS.506.2603M}, the \acp{IC} were generated at the starting redshift $z = 1/a - 1 = 127$ using the {\NGenIC} code \citep{2015ascl.soft02003S}, which employs the Zel'dovich approximation (or second-order Lagrangian perturbation theory) to generate a random realization of density fluctuations consistent with a prescribed power spectrum in terms of a perturbed, but otherwise regular particle distribution.
We analyse and compare four different kinds of simulations (cf.\ \cref{sec:simulations}), differentiated by the solved dynamics (\ac{FDM}/Schrödinger–Poisson and \ac{CDM}/$N$-body) and the initial power spectrum (\q{standard} \ac{LCDM} and \q{self-consistent} \ac{FDM} \acp{IC} with a cut-off at small scales).
For each case, a different approach for the \acp{IC} was necessary:
\begin{itemize}
	\item \ac{CDM} ($N$-body) dynamics:
	In this case, the particle distribution generated by {\NGenIC} can simply be used directly.
	\item \ac{FDM} (Schrödinger–Poisson) dynamics:
	The initial wave function is constructed using the prescription
	\begin{align}
		\label{eq:ic-absolute}
		|ψ(\vec{x})| &= √{\frac{ρ(\vec{x})}{m}},
		\displaybreak[0]
		\\
		\label{eq:ic-phase}
		∇ \arg\ml(ψ(\vec{x})\mr)
		&= ∇ θ(\vec{x})
		= \frac{m}{ℏ} \vec{v}(\vec{x}),
	\end{align}
	where $ρ$ is the matter density and $\vec{v}$ the velocity \citep[cf.\ \cref{eq:mass-density,eq:velocity};][]{1993ApJ...416L..71W, 2018PhRvD..97h3519M, 2021MNRAS.506.2603M}.
	In this case, the random realization of the density fluctuation $δ(\vec{x})$ in {\NGenIC} is actually used without creating a particle distribution at all, in order to avoid unnecessary transformation steps.
	For the absolute value of the complex wave function, the correspondence to the density via \cref{eq:ic-absolute} is straightforward.
	The wave function's phase requires a few more steps, with the following result using \cref{eq:ic-phase} (in Fourier space)
	\begin{equation}
		ℱ[θ](\vec{k})
		= -\imag \frac{m}{ℏ} \frac{\vec{k}}{k^2} ⋅ \FFT[\vec{v}](\vec{k})
	\end{equation}
	where $\vec{v}$ is the velocity field in the Zel'dovich approximation.
	\item \ac{CDM} initial power spectrum:
	The simulations for this case are the ones previously presented in \citet{2021MNRAS.506.2603M}.
	The \acp{IC} were generated with an input power spectrum following \citet{1990Natur.348..705E,1992MNRAS.258P...1E}, i.\,e.\ of the form
	\begin{equation}
		\label{eq:efstathiou}
		P_{\mtext{CDM}}(k) ∝
		k \ml[1 + \ml(ak + (bk)^{\frac{3}{2}} + c^2 k^2\mr)^ν\mr]^{-\frac{2}{ν}} ,
	\end{equation}
	with $a = \num{6.4}/Γ\,\si{\per\hHubble\Mpc}$, $b = \num{3.0}/Γ\,\si{\per\hHubble\Mpc}$, $c = \num{1.7}/Γ\,\si{\per\hHubble\Mpc}$, $Γ = Ω_{\mtext{m}}h = \num{0.21}$,\footnote{%
		See \cref{sec:simulations}.%
	} and $ν = \num{1.13}$. Here $h = H_0 / (\SI{100}{\km\per\s\per\Mpc})$ encodes the Hubble constant.
	\item \ac{FDM} initial power spectrum:
	In this case, we make use of the power spectrum predicted in linear theory by the code \textcode{axionCAMB} \citep{2015PhRvD..91j3512H,2022ascl.soft03026G} for an axion cosmology.
	However, in order to be able to compare our simulations as directly as possible, we aim to minimize differences in the \ac{CDM} and \ac{FDM} \acp{IC} arising from the two different methods (a fitting function \cref{eq:efstathiou} and a numerical Boltzmann code).
	To this end, instead of using the power spectrum $P_{\text{\textcode{axionCAMB},FDM}}(k)$ from \textcode{axionCAMB} directly, we also calculated the corresponding \ac{LCDM} power spectrum $P_{\text{\textcode{axionCAMB},CDM}}(k)$ using the same code and obtained the final input spectrum $P_{\mtext{FDM}}(k)$ as follows:
	\begin{equation}
		P_{\mtext{FDM}}(k) = \frac{P_{\text{\textcode{axionCAMB},FDM}}(k)}{P_{\text{\textcode{axionCAMB},CDM}}(k)} P_{\mtext{CDM}}(k).
	\end{equation}
	This ensures that any variations between \cref{eq:efstathiou} and \textcode{axionCAMB} are cancelled out, and only the differences arising from the choice of the \ac{CDM} or \ac{FDM} model remain.
\end{itemize}

\subsection{Simulations}
\label{sec:simulations}

\begin{table}
	\centering
	{%
	\sisetup{list-pair-separator={, }, list-final-separator={, }}%
	\setlength{\tabcolsep}{5pt}%
	\begin{tabular}{
		l l @{\,} r @{\enspace} S @{\,} c S
	}
		\toprule
		Type &
		IC &
		Res.\ el. &
		{$L$ / \si{\per\hHubble\Mpc}} &
		{$m c^2$ / \si{\eV}} &
		{Resolution}
		\\
		\midrule
		FDM & FDM & $8640^3$ & 10 & \hfill \num{7e-23} & \SI{1.16}{\per\hHubble\kpc}
		\\
		FDM & FDM & $6144^3$ & 10 & \hfill \num{5e-23} & \SI{1.63}{\per\hHubble\kpc}
		\\
		FDM & FDM & $4320^3$ & 10 & \hfill \num{3.5e-23} & \SI{2.31}{\per\hHubble\kpc}
		\\
		FDM & FDM & $4320^3$ & 10 & \hfill \num{2e-23} & \SI{2.31}{\per\hHubble\kpc}
		\\
		FDM & FDM & $4320^3$ & 10 & \hfill \num{1e-23} & \SI{2.31}{\per\hHubble\kpc}
		\\[\smallskipamount]
		FDM & CDM & $8640^3$ & 10 & \hfill \num{7e-23} & \SI{1.16}{\per\hHubble\kpc}
		\\
		FDM & CDM & $4320^3$ & 10 & \hfill \num{3.5e-23} & \SI{2.31}{\per\hHubble\kpc}
		\\[\smallskipamount]
		CDM & FDM & $2048^3$ & 10 & \hfill \num{7e-23} & \SI{9.69e3}{\per\hHubble\Ms}
		\\
		CDM & FDM & $2048^3$ & 10 & \hfill \num{3.5e-23} & \SI{9.69e3}{\per\hHubble\Ms}
		\\
		CDM & CDM & $2048^3$ & 10 & {—} & \SI{9.69e3}{\per\hHubble\Ms}
		\\
		\bottomrule
	\end{tabular}%
	}
	\caption{%
		List of performed simulations with important characteristics: simulation type (\acs*{FDM}/\acs*{SP} or \acs*{CDM}/$N$-body), number of resolution elements (grid cells or $N$-body particles), box size, \acs*{FDM} particle mass, and resolution (grid cell size or $N$-body particle mass).
		The lengths given for the box sizes and resolutions are comoving.
		The simulations with \acs*{CDM} \acsp*{IC} have previously been presented in \citet{2021MNRAS.506.2603M}.
	}
	\label{tab:simulations}
\end{table}

\begin{figure*}
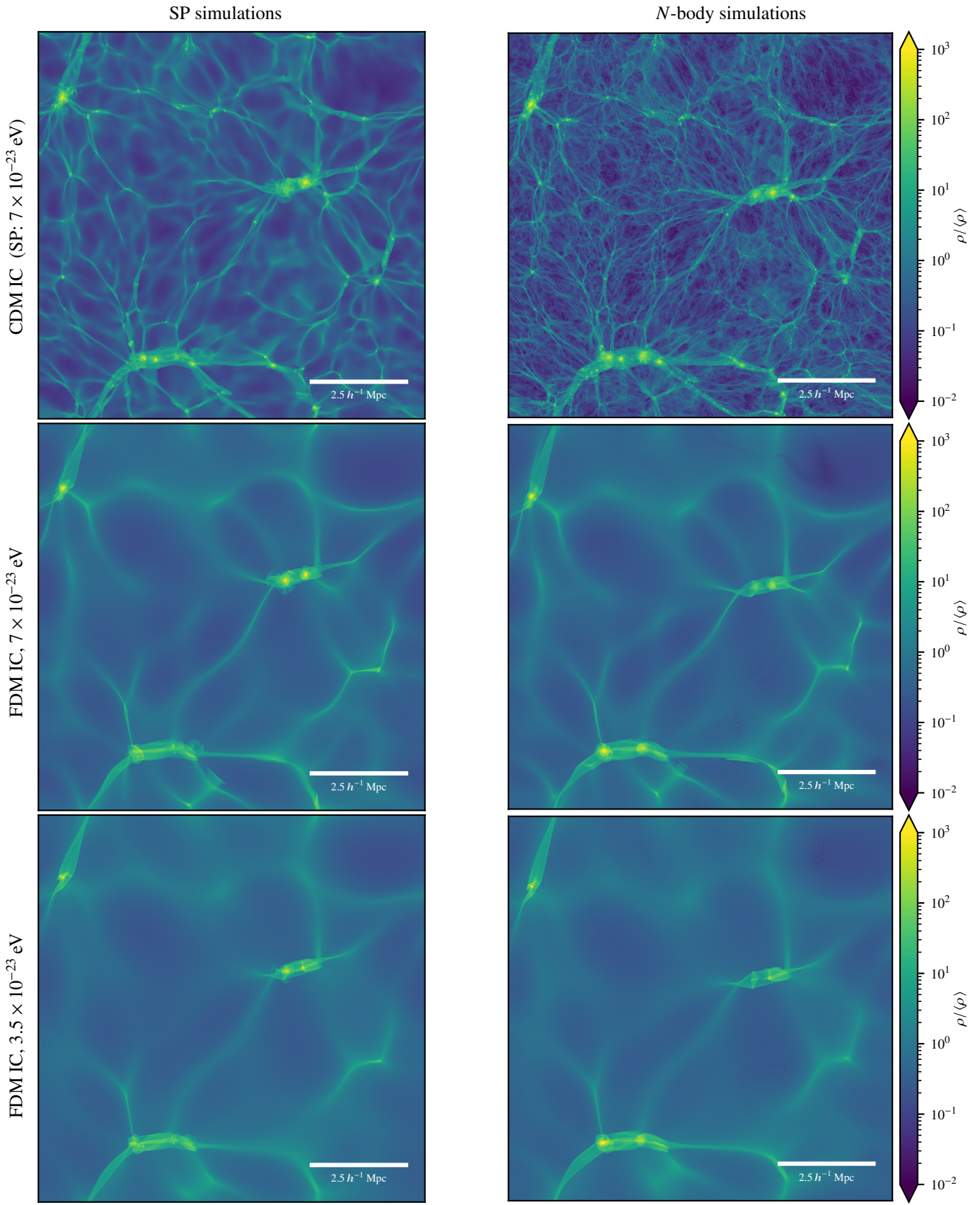

	\centering
	\large

	\hspace*{0.35\columnwidth}%
	\ac{SP} simulations%
	\hfill%
	$N$-body simulations%
	\hspace*{0.35\columnwidth}%

	\smallskip

	\rotatebox{90}{\strut\hspace{4.5em}\smash{CDM IC \;(\ac{SP}: \SI{7e-23}{\eV})}}%
	\hspace{1ex}%
	\import{res/}{project_fdm_cdmic_7e-23.pgf}%
	\hfill%
	\import{res/}{project_cdm_cdmic.pgf}

	\rotatebox{90}{\strut\hspace{5em}\smash{FDM IC, \SI{7e-23}{\eV}}}%
	\hspace{1ex}%
	\import{res/}{project_fdm_axionic_7e-23.pgf}%
	\hfill%
	\import{res/}{project_cdm_axionic_7e-23.pgf}

	\rotatebox{90}{\strut\hspace{4.5em}\smash{FDM IC, \SI{3.5e-23}{\eV}}}%
	\hspace{1ex}%
	\import{res/}{project_fdm_axionic_3.5e-23.pgf}%
	\hfill%
	\import{res/}{project_cdm_axionic_3.5e-23.pgf}

	\caption{%
		Projected dark matter densities along a thin (\SI{100}{\per\hHubble\kpc}) slice in cosmological box simulations for different dynamics (\ac{FDM}/\ac{SP}, left column; or \ac{CDM}/$N$-body, right column) and \acp{IC} (\ac{FDM} or \ac{CDM}, rows), for box sizes $L = \SI{10}{\per\hHubble\Mpc}$ at $z = 3$.%
	}
	\label{fig:projections-overview}
\end{figure*}

The simulations were performed using the cosmological parameters $Ω_{\mtext{m}} = \num{0.3}$, $Ω_{\mtext{b}} = \num{0}$, $Ω_Λ = \num{0.7}$, $H_0 = \SI{70}{\km\per\s\per\Mpc}$ ($h = \num{0.7}$), and $σ_8 = \num{0.9}$,\footnote{%
	As usual, the density parameters $Ω_i$ indicate the values at $z = 0$.%
}
with \acp{IC} as described in \cref{sec:ics}.
The same cosmological parameters and random seeds were shared for the generation of all \acp{IC} at $z = 127$ in order to allow for a direct comparison between different dark matter models and resolutions, including those previously presented in \citet{2021MNRAS.506.2603M}.
For a comoving box size of \SI{10}{\per\hHubble\Mpc}, and different masses $m$, simulations with different resolutions – up to grid sizes of $N^3 = 8640^3$ – were performed.
As before, the simulations were run until $z = 3$, where the \ac{FDM} simulations should still be largely unaffected by resolution effects.
Furthermore, even the largest modes in simulations of this box size would become non-linear before $z = 0$, making the reliability and benefit of evolving simulations to this point dubious in any case.
A detailed list of the different simulations is given in \cref{tab:simulations}.
A visual overview showing the projected density across a slab of the simulation volume is displayed for a subset of the simulations in \cref{fig:projections-overview}.
The images readily make differences between the four different kinds of simulations we have carried out apparent, and they show the impact of different values of $m$ as well.

Most of the simulations listed in \cref{tab:simulations} focus on the mass values $mc^2 = \SIlist{3.5e-23; 7e-23}{\eV}$, respectively, used in \citet{2021MNRAS.506.2603M}, but we also performed additional \q{self-consistent} \ac{FDM} simulations (\ac{SP} + \ac{FDM} \acp{IC}) for $mc^2 = \mbox{\SIlist{1e-23; 2e-23; 5e-23}{\eV}}$ in order to investigate the dependence of \ac{FDM} phenomenology on the particle mass.
For the lighter masses \SIlist{1e-23; 2e-23}, the resolution (with respect to \cref{eq:velocity-criterion-x}) is actually much better than that of the other simulations, allowing us to evolve them further in time beyond $z = 3$ (although once again with the caveat that all spatial modes eventually become non-linear).
For $N$-body simulations, a very high mass resolution (large number of particles) was used when compared to most similar cosmological simulations.
This was done in order to be able to measure the power spectrum down to very small scales, comparable to those accessible using a $8640^3$ grid.
Due to our earlier work in \citet{2021MNRAS.506.2603M} using the same parameters, it was not necessary to investigate the numerical convergence of the simulations again in detail.

Although progress has been made e.\,g.\ in hybrid methods \citep{2022PhRvL.128r1301S}, which do not solve the equations of motion for the full wave dynamics everywhere, our simulations remain the largest cosmological \ac{SP} simulations of structure formation with \ac{FDM}.
The new simulations with \ac{FDM} \acp{IC} allow us to perform a four-way comparison between \ac{SP} and $N$-body dynamics on the one hand, and \ac{FDM} and \ac{CDM} \acp{IC} on the other hand, and to quantify the impact of both aspects.
Apart from being able to study a self-consistent \ac{FDM} cosmology, this also allows us to gain insight into the extent to which \ac{FDM} can be approximated (e.\,g.) by using $N$-body simulations with a modified initial power spectrum, and to examine the validity of previous work which took this approach \citep[e.\,g.][]{2016ApJ...818...89S}.

\begin{figure*}
	\centering

	\import{res/}{project_mark_fof60.pgf}%
	\hfill%
	\raisebox{1mm}{\includegraphics[width=0.45\textwidth]{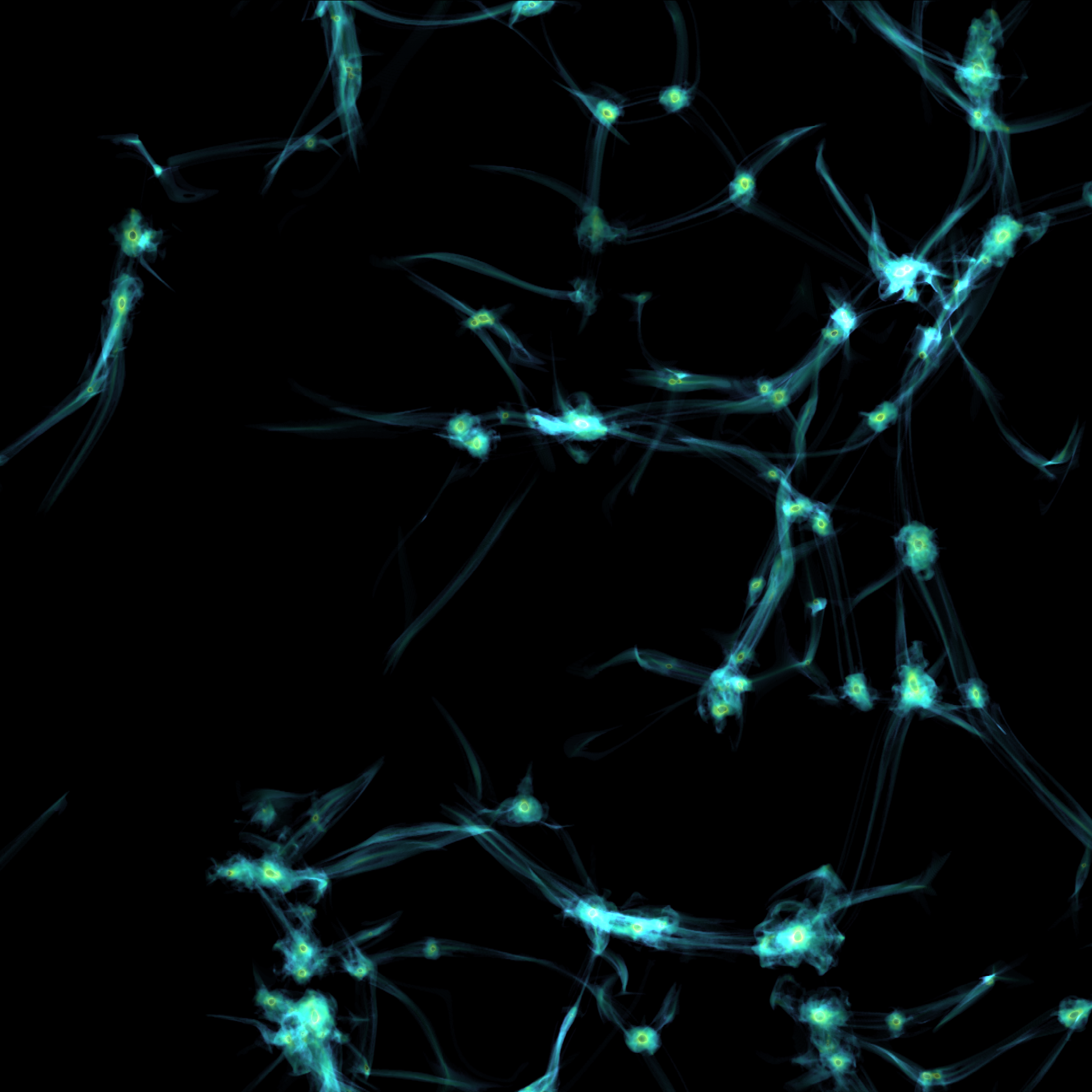}}%

	\import{res/}{project_mark_fof100.pgf}%
	\hfill%
	\import{res/}{project_mark_fof200.pgf}%
	\hfill%
	\raisebox{-1.6mm}{\includegraphics[scale=1.03]{project_mark_fof300.pgf}}%

	\caption{%
		Projected dark matter density and volume rendering (different viewing angle) showing the largest \acs*{FoF} group using a halo finder overdensity threshold of $60ρ_{\mtext{m}}$ (top row), and $100ρ_{\mtext{m}}$, $200ρ_{\mtext{m}}$ and $300ρ_{\mtext{m}}$ (bottom row) in a cosmological box simulation of \acs*{FDM} at $z = 3$ with box size $L = \SI{10}{\per\hHubble\Mpc}$, grid size $N^3 = 8640^3$, \acs*{FDM} mass $m c^2 = \SI{7e-23}{\eV}$, and \acs*{FDM} \acsp*{IC}.
		The areas marked in red/orange in the projections and shown in the volume rendering indicate regions spanned by the largest \acs*{FoF} group identified by the halo finder using the given overdensity threshold.%
	}
	\label{fig:mark-fof}
\end{figure*}

\subsection{Halo identification with suppressed small-scale power}
\label{sec:halo-finding}

In order to identify dark matter haloes, the \ac{FoF}-like halo finder developed in \citet{2021MNRAS.506.2603M}, which is able to work on a Cartesian grid, was used.
Instead of a linking length for particle distances, this grid-based halo finder uses a density threshold as an analogous parameter.
Only grid cells above the given overdensity threshold are considered, and they are always linked if they are adjacent.

\begin{figure}
	\centering
	\import{res/}{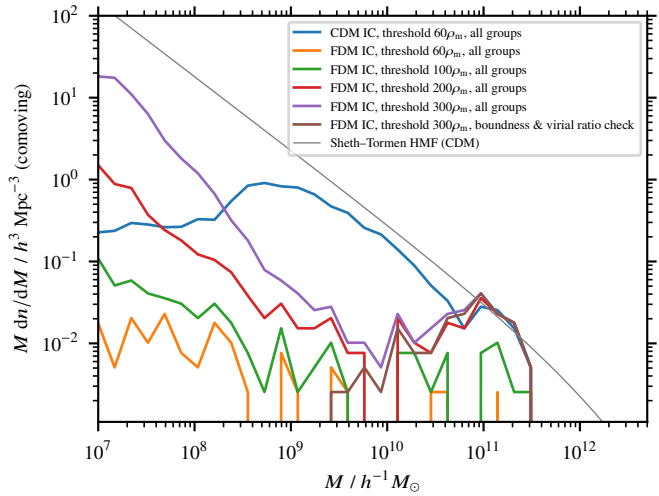}

	\caption{%
		Halo mass function of a cosmological \acs*{FDM} simulations at $z = 3$ with a box size of $L = \SI{10}{\per\hHubble\Mpc}$ and \acs*{FDM} \acsp*{IC} for different values of the halo finder's overdensity threshold and after filtering using the gravitational binding criterion. Different simulations simulations are shown as labelled.%
	}
	\label{fig:halo-finder-thresholds}
\end{figure}

While this approach worked very well for \ac{SP} simulations with \ac{CDM}-like \acp{IC} and an overdensity threshold of $60ρ_{\mtext{m}}$, the results with \ac{FDM} \acp{IC} were very different.
In this case, haloes are linked via \emph{continuous, smooth, dense filaments} throughout the entire simulation volume.
The central densities of these filaments exceed the threshold value of $60ρ_{\mtext{m}}$, leading to much larger regions of space being linked by the halo finder than intended.
Indeed, as illustrated in \cref{fig:mark-fof}, the largest \ac{FoF} group traces a network of filaments and spreads across the entire simulation box, incorporating numerous haloes and filaments.
This has dire consequences for the endeavour of actually identifying individual haloes:
Because each \ac{FoF} group is counted at most as a single halo, the largest group subsumes many haloes and thus leads to a severe under-counting.
As shown in \cref{fig:halo-finder-thresholds}, the resulting \ac{HMF} shows very few or even no haloes at all across wide mass ranges.

\begin{figure}
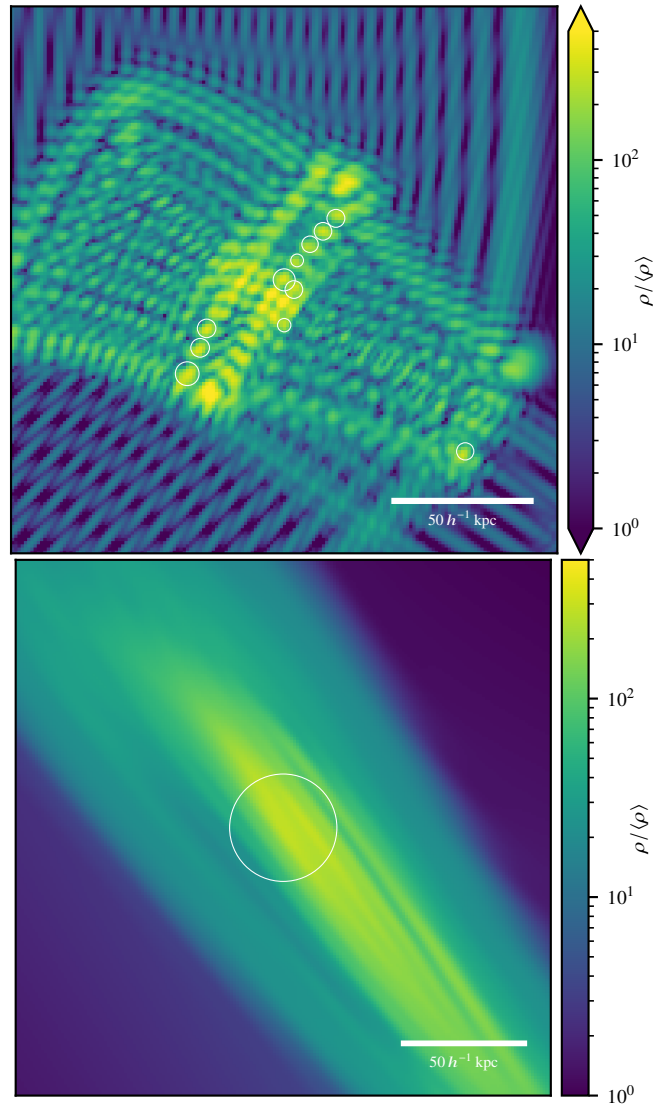

	\centering

	\import{res/}{misidentified_interference.pgf}

	\import{res/}{misidentified_filament.pgf}

	\caption{%
		Examples of spurious \ac{FoF} groups identified in \ac{FDM} interference patterns (top) and filaments (bottom) in a cosmological \ac{FDM} simulation (projected density).
		The former arise in great numbers as a consequence of raising the halo finder's overdensity threshold to $300ρ_{\mtext{m}}$, while the latter appear due to a lack of small-scale power analogous to \ac{WDM}.
		The circles mark the groups' locations, with the radius corresponding to $R_{200}$.%
	}
	\label{fig:false-haloes}
\end{figure}

\begin{figure}
	\centering

	\import{res/}{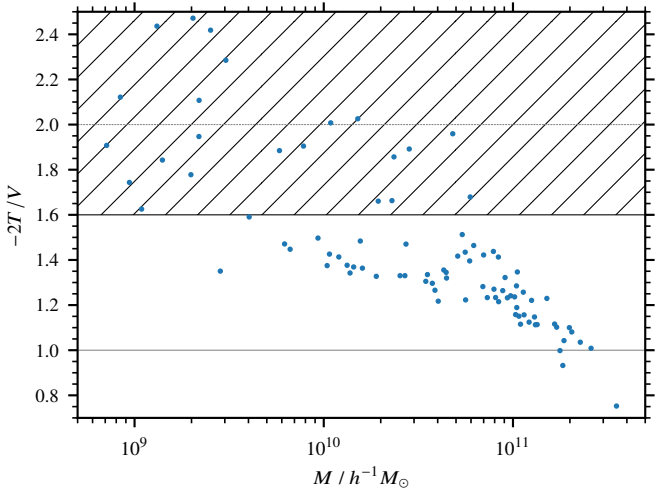}

	\caption{%
		Virial ratios $-2T / V$ for \ac{FoF} groups identified by {\AxiREPO}'s halo finder with an overdensity threshold of $300ρ_{\mtext{m}}$ in a cosmological \ac{FDM} simulation at $z = 3$ with $L = \SI{10}{\per\hHubble\Mpc}$, $N^3 = 8640^3$, $m c^2 = \SI{7e-23}{\eV}$ and \ac{FDM} \acp{IC}.
		The potential energy $V$ is the \q{self-potential}, and both $T$ and $V$ are summed within a radius of $R_{200}$ for each halo.
		Points above the dashed line correspond to \ac{FoF} groups which are not gravitationally bound ($E = T + V > 0$).
		Points in the hatched region ($-2T / V > \num{1.6}$) are excluded from our definition of \q{haloes}.%
	}
	\label{fig:virial-ratios}
\end{figure}

In order to break the filamentary links between haloes, it was necessary to use a higher value for the overdensity threshold.
Although the \ac{FoF} groups are smaller as a consequence, encompassing only the denser inner parts of haloes, this removes the filaments, which do not typically reach such high densities, from the groups.
Empirically, we arrived at using a value of $300ρ_{\mtext{m}}$ to reliably prevent spurious connections between and merging of \ac{FoF} groups (cf.\ \cref{fig:halo-finder-thresholds}).

However, this countermeasure invokes another problematic effect:
Now, \ac{FoF} groups are broken up into many small parts which do not all correspond to actual and reasonably complete haloes.
\Cref{fig:halo-finder-thresholds} shows how increasing the group finder's threshold leads to an ever larger proliferation of low-mass objects in the \ac{HMF}.
Clearly, these objects are spurious artefacts, since, as visible in \cref{fig:halo-finder-thresholds}, their number keeps increasing with the threshold parameter, eventually even exceeding the \ac{CDM} \ac{IC} case, which can be viewed as an upper limit for the \ac{HMF} expected for \ac{FDM} \acp{IC}.

On visual inspection of a sample of these objects, it turned out that the vast majority of them are localized density fluctuations arising from the wave nature of \ac{FDM}, e.\,g.\ in the form of constructive interference fringes, which can reach very high densities.
The top panel of \cref{fig:false-haloes} shows how a large number of transient interference maxima is erroneously identified as haloes.
Although the same patterns are present in the \ac{FDM} simulations with \ac{CDM} \acp{IC}, the lower threshold value of $60ρ_{\mtext{m}}$ ensured that these objects remained connected to haloes within a single \ac{FoF} group.

In the end, filtering this set of \ac{FoF} groups using a gravitational binding criterion proved necessary and successful for reliably identifying haloes.
To this end, we compute the \q{self-potential} of each \ac{FoF} group within $R_{200}$, i.\,e.\ the gravitational potential $V_{\mtext{s}}$ generated only by the matter within $R_{200}$,\footnote{%
	The self-potential calculation was here done using direct summation, approximating each grid cell as a point particle for simplicity, which is accurate up to hexadecupole order in the multipole expansion \citep{Barnes1989}.%
}
\begin{equation}
	V_{\mtext{s}}
	= -\frac{1}{2} ∑_{\substack{i, j\\i ≠ j}}^{|\vec{r}_i - \vec{r}_j| < R_{200}} \frac{G m_i m_j}{|\vec{r}_i - \vec{r}_j|}
	,
\end{equation}
as well as the kinetic energy $T_v$ in the group's center-of-mass frame using \cref{eq:momentum-density} and the gradient energy $T_ρ$, and excluded any \ac{FoF} group which did not meet the binding criterion
\begin{equation}
	\label{eq:bound-criterion}
	E = T_v + T_ρ + V_{\mtext{s}} < 0
	,
\end{equation}
cf.\ \cref{eq:energy}.
This reliably excludes the gravitationally unbound, transient wave interference patterns.

It should be noted that the computation of velocities from the complex wave amplitude $ψ$ using the gradient of the phase as in \cref{eq:velocity} is not straightforward.
Not only is it the phase a periodic variable, making it necessary to take into account wrap-arounds at $2π$ when computing gradients, but it is also undefined in regions of destructive interference, where $ρ → 0$.
In the end, we found the most robust method to be using the (equivalent) definition via the momentum density in \cref{eq:momentum-density}.
Not only does this avoid the treatment of the periodic phase variable, but it is also well-defined even for $ρ = 0$, preventing pathological cases (extremely large velocities) when computing the numerical derivative.
Although obtaining the velocity from the momentum density requires dividing by $ρ$, the results of this procedure turn out to be much more well-behaved in practice.

Finally, in addition to the above, we enforced a stricter cut on the virial ratio $-2T / V$ in the form of
\begin{equation}
	\label{eq:virial-ratio-cut}
	-2T / V_{\mtext{s}} < 1.6
\end{equation}
in order to eliminate a number of objects which were clearly highly perturbed objects and not reasonably relaxed haloes.\footnote{%
	Since \cref{eq:bound-criterion} corresponds to $-2T / V_{\mtext{s}} < 2$, \cref{eq:virial-ratio-cut} automatically enforces gravitational binding as well.%
      }
Similar cuts are regularly applied in \ac{CDM} models when density profiles of haloes are studied \citep[e.\,g.][]{2007MNRAS.381.1450N}.
These objects are distinct from the wave interference phenomena, and instead could be called very high-density filamentary structures, arising from the lack of power on small scales in the \ac{FDM} \acp{IC} case.
Their high virial ratios indicate that, while gravitationally bound, they have not reached a virialised state.
In contrast, for \ac{CDM}, such filaments do not remain smooth, but fragment into (sub-)haloes down to the smallest scales.
Examples of both kinds of non-halo objects are shown in \cref{fig:false-haloes}.

\Cref{fig:virial-ratios} shows the virial ratios for \ac{FoF} groups in an \ac{FDM} simulation as a function of $M_{200}$ (determined using a spherical overdensity algorithm).
Although a virial ratio of $≈ 1$, corresponding to $2T ≈ |V|$, would be expected for virialised objects, there is a significant bias of many objects around $-2T / V_{\mtext{s}} ≈ \num{1.2}$.
Indeed, this effect has been observed for \ac{CDM} haloes in $N$-body simulations as well \citep{2007MNRAS.376..215B, 2007MNRAS.381.1450N}, and seems to be due to the definition of the self-potential, which neglects the contribution from a possible large-scale tidal field \citep{2021MNRAS.508.5196S}.
A more sophisticated approach, such as the \q{boosted potential binding check} from \citet{2021MNRAS.508.5196S}, can potentially improve the results, but was not necessary in our case, perhaps due to the low number of remaining haloes after selection, which allows for easy manual inspection of the end result.
Notably, these problems of
\begin{inparaenum}[(1)]
	\item enormous \ac{FoF} groups, linked by smooth filaments, stretching across 10s of \si{\Mpc}, and
	\item additional, non-virialised structures identified by the halo finder,
\end{inparaenum}
have been encountered in $N$-body simulations of \ac{WDM} before \citep[esp.\ figs.\ 2 and 4]{2013MNRAS.434.3337A, 2021MNRAS.508.5196S}. There have also been speculations that stars themselves may form in \ac{WDM} filaments \citep{Gao2007}, giving rise to a qualitatively different mode of galaxy formation compared to \ac{CDM}.

\begin{figure*}
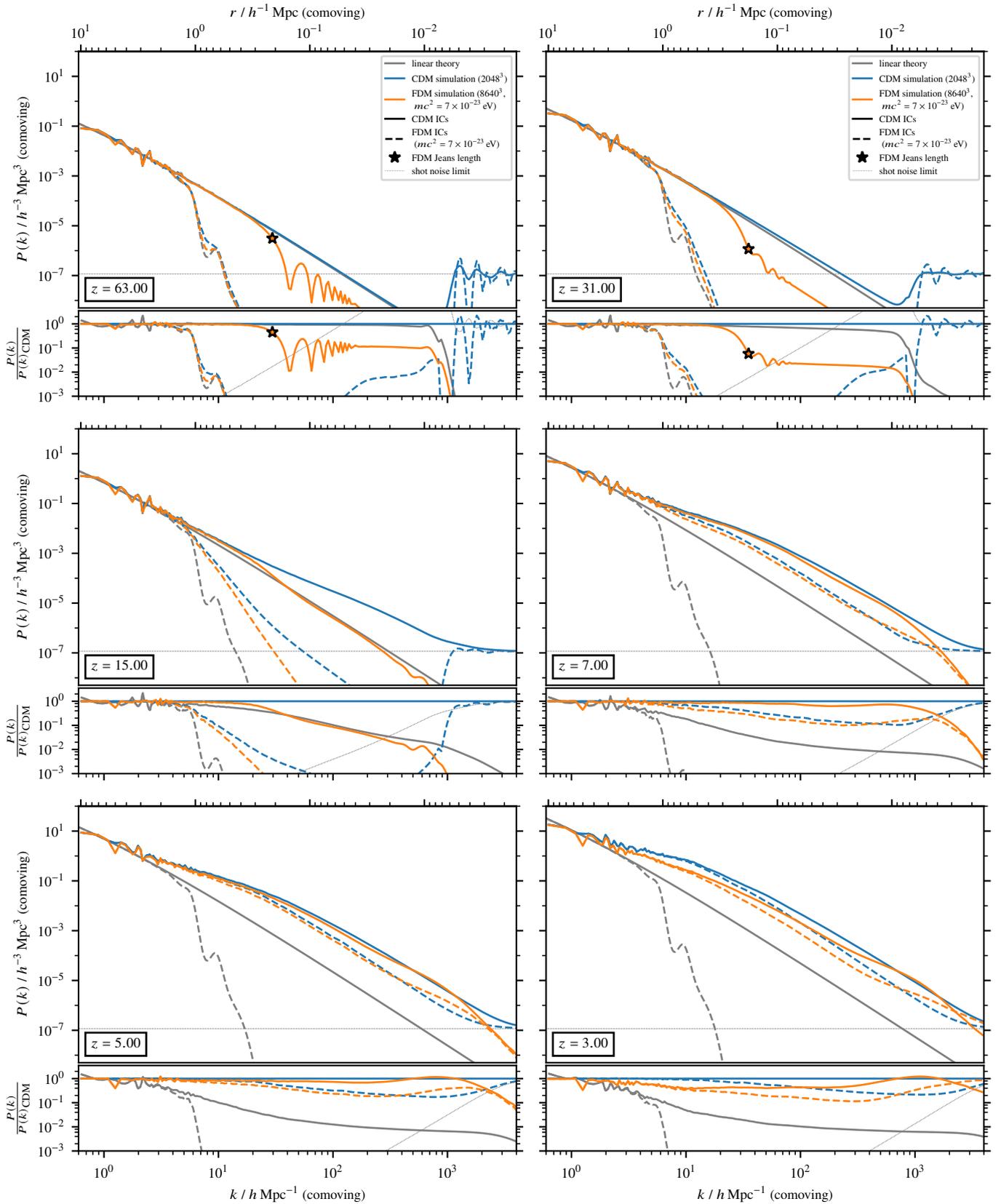

	\centering

	\import{res/}{powerspec_0.0156.pgf}%
	\hfill%
	\hspace*{-0.4in}%
	\import{res/}{powerspec_0.0312.pgf}

	\vspace*{-10ex}

	\import{res/}{powerspec_0.0625.pgf}%
	\hfill%
	\hspace*{-0.4in}%
	\import{res/}{powerspec_0.1250.pgf}

	\vspace*{-10ex}

	\import{res/}{powerspec_0.1667.pgf}%
	\hfill%
	\hspace*{-0.4in}%
	\import{res/}{powerspec_0.2500.pgf}

	\caption{%
		Four-way comparison of dark matter power spectra at different redshifts for cosmological \acs*{FDM} (wave) and \acs*{CDM} ($N$-body) simulations with \acs*{FDM} and \acs*{CDM} \acp*{IC} in $L = \SI{10}{\per\hHubble\Mpc}$ boxes.
        The power spectrum evolved using linear perturbation theory is shown for comparison.
      	The lower panels show the ratio of the power spectra to the \acs*{CDM} result ($N$-body simulation with \acs*{CDM} \acp*{IC}).
      	For $z = 63$, the dashed line additionally indicates the \acs*{FDM} Jeans scale (\cref{eq:jeans}).
      	Faint dotted lines show the shot noise limits of the $N$-body simulations; the power spectrum cannot be measured accurately once this limit is reached.%
	}
	\label{fig:power-spectra-fdm-vs-cdm}
\end{figure*}

\section{Four-way comparison of the power spectrum}
\label{sec:power-spectrum}

The matter power spectrum is a crucial measure of matter clustering at different length scales.
For \acs{LCDM}, even including baryonic effects, its behaviour and temporal evolution have been determined rather accurately using numerical simulations \citep[e.\,g.][]{1998ApJ...499...20J, Hellwing2016, 2018MNRAS.475..676S}.
For \ac{FDM}, linear structure formation is scale-dependent even in linear perturbation theory, with a suppression at scales smaller than the \ac{FDM} Jeans length $λ_{\mtext{J}} = 2π / k_{\mtext{J}}$ \eqref{eq:jeans}, and even during non-linear evolution on scales where the wave nature of \ac{FDM} is relevant, i.\,e.\ at or below the de Broglie wavelength $λ_{\mtext{dB}}$ \eqref{eq:de-broglie}.
Intuitively, this can be understood as a consequence of the analogue of the Heisenberg uncertainty principle (Schrödinger formulation), or, equivalently, the presence of the so-called \q{quantum pressure} (Madelung fluid formulation).

Combined with our previous results \citep{2021MNRAS.506.2603M}, our new simulations with \ac{FDM} \acp{IC}, using both \ac{SP} and $N$-body solvers, allow us to draw a four-way comparison of observables, contrasting results for each combination of \acp{IC} (\ac{FDM} vs.\ \ac{CDM}) and dynamics (\ac{SP}/\q{\ac{FDM}} vs.\ $N$-body/\q{\ac{CDM}}).
Although only two of the four cases are strictly self-consistent in a physical sense (namely, \ac{FDM} and \ac{CDM} with their corresponding \acp{IC}), this approach allows us to disentangle the two essential physical differences distinguishing \ac{FDM} from \ac{CDM} in cosmological numerical simulations: the initial conditions and the dynamics on small scales.
This is particularly relevant to evaluate to what extent approximate methods, such as the use of $N$-body simulations with an \ac{FDM} initial power spectrum \citep{2016ApJ...818...89S,2020MNRAS.494.2027M}, can yield reasonable results.

In \Cref{fig:power-spectra-fdm-vs-cdm}, we show the measured matter power spectra for both \ac{FDM} and \ac{CDM} cases, where simulations with $mc^2 = \SI{7e-23}{\eV}$ are used to represent \ac{FDM}.
The solid lines (\ac{CDM} \acp{IC}) correspond to the results from \citet{2021MNRAS.506.2603M}, whereas the dashed lines are the new simulations with \ac{FDM} \acp{IC} added in this work.
Reassuringly, all cases accurately agree with each other and with linear perturbation theory on large scales and at early times, and it can be observed again that the onset of non-linear structure formation is delayed for \ac{FDM}, and does not proceed strictly in the same bottom-up
fashion as for \ac{CDM}.

As a reflection of this fact, the differences between the four cases exhibit variations across time.
For example, only the $N$-body \ac{CDM} simulation significantly exceeds linear growth at redshifts below $15$, although the \ac{SP} \ac{CDM} case has similar, mild non-linear amplification on large enough scales of $k ≈ \mbox{\SIrange{10}{20}{\hHubble\per\Mpc}}$ at $z = 15$.
However, by $z = 7$, considerable non-linear enhancement is present on small scales $k ≳ \SI{5}{\hHubble\per\Mpc}$ in all cases.

Indeed, for $z ≥ 7$, the power spectra for simulations with the same \acp{IC} track each other comparatively closely on all scales, although the \ac{SP} simulations are still suppressed by some tens of percent for $k ≳ \SI{10}{\hHubble\per\Mpc}$.
There seems to be little qualitative change in the relative evolutions of the power spectra beyond $z = 7$.
Notably, for the most part during this time, the difference between \ac{SP} and $N$-body is significantly smaller than that between the different sets of \acp{IC}, and is of similar size in both cases.

At $z = 3$, \ac{SP} and $N$-body results drift more apart compared to earlier times ($z = 7$, $z = 5$).
Interestingly, at this time, the relative difference between the dynamics and the \acp{IC} is now roughly the same, meaning that each of the two physical \q{ingredients} has a similar impact at this time.
Since the time evolution stops at $z = 3$, it is however unclear whether this is the onset of a new, sustained phenomenological difference or whether it is a transient, numerical artefact.
Since the resolution requirements for \ac{SP} grow more stringent towards later times, it is also possible that these slight relative changes in the power spectra are the first hints of resolution effects affecting the results.

An interesting difference for the \ac{SP} simulations with different \acp{IC} is the location of the \q{bump} on small scales, which starts to appear at $z = 7$.
At the location of this bump, the small-scale power in \ac{SP} simulations matches or even slightly exceeds the \ac{CDM} power (for \ac{FDM} \acp{IC}, the \ac{SP} simulation quite significantly surpasses its $N$-body companion on these scales).
It is commonly attributed to the presence of \ac{FDM} interference patterns and granules of size $λ_{\mtext{dB}}$ \citep{2020MNRAS.494.2027M}, which are unique to the wave nature of \ac{FDM} and not present in \ac{CDM}.
However, curiously, there is a shift in the location of this bump for the two \ac{SP} simulations, with the bump in the \ac{FDM} \ac{IC} case appearing on smaller scales.
Following the interpretation that the bump is linked to $λ_{\mtext{dB}}$, this would imply that the \ac{FDM} simulation displays more power on smaller wavelengths, which (with the other parameters being equal) would correspond to higher velocities.
The physical origin of this effect remains unclear and is a subject for future investigation.

\section{The halo mass function}
\label{sec:halo-mass-function}

\begin{figure*}
	\centering

	\input{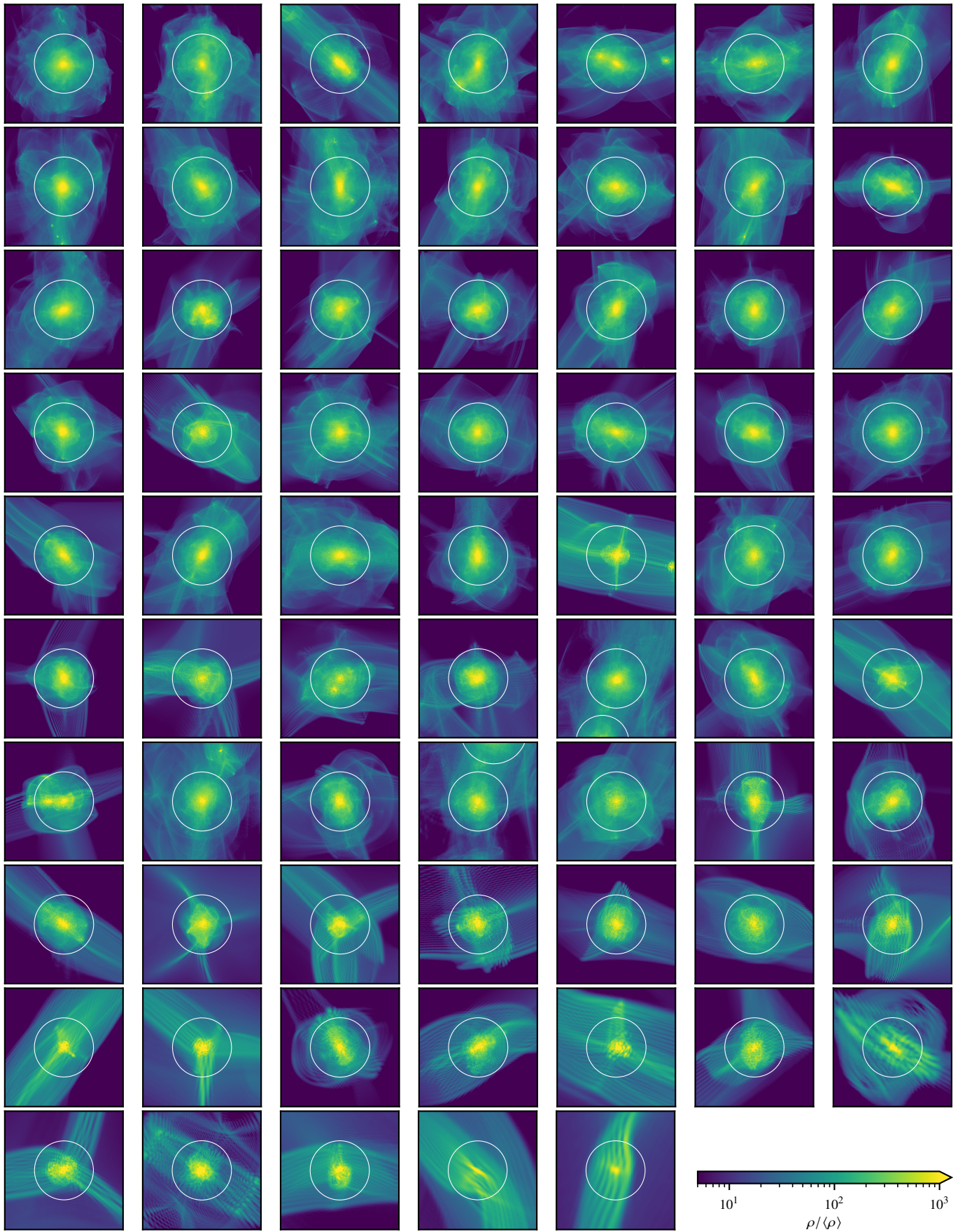}

	\caption{%
		Projected dark matter densities for all 68 identified haloes at $z = 3$ in a cosmological \ac{FDM} simulation with box size $L = \SI{10}{\per\hHubble\Mpc}$, grid size $N^3 = 8640^3$, \acs*{FDM} mass $m c^2 = \SI{7e-23}{\eV}$, and \acs*{FDM} \acsp*{IC}.
		The projection depth is $2R_{200}$ for each halo.%
	}
	\label{fig:halo-projections}
\end{figure*}

\begin{figure}
	\centering
	\import{res/}{mass_function_7e-23.pgf}

%

	\caption{%
		Halo mass functions of cosmological \acs*{FDM} and \acs*{CDM} simulations at $z = 3$ with \ac{FDM} mass $mc^2 = \SI{7e-23}{\eV}$, a box size of $L = \SI{10}{\per\hHubble\Mpc}$ and different \acsp*{IC}.
		The \acs*{HMF} derived for \acs*{CDM} by \citet{1999MNRAS.308..119S} is shown for comparison.
		Different predictions for the \ac{FDM} \ac{HMF} from \citet{2016ApJ...818...89S, 2017MNRAS.465..941D, 2022MNRAS.510.1425K} are shown using dash-dotted lines.
		The lower panel shows the ratios of the mass functions to the result of the \acs*{CDM} simulation.%
	}
	\label{fig:halo-mass-function}
\end{figure}

The formation of gravitationally collapsed structures of dark matter, so-called haloes, is one of the most significant results of cosmic structure.
In particular, haloes serve as the loci of collapse of baryonic matter and thus provide the environment for the formation and evolution of galaxies.
The \ac{HMF}, which is a measure of halo abundance as a function of mass, and its evolution across time, are thus critical benchmarks of a cosmological model.

While the (extended) Press–Schechter formalism \citep{1974ApJ...187..425P, 1999MNRAS.308..119S} has been demonstrated to be a simple and reliable method to estimate the \ac{HMF} and its evolution (depending only on the power spectrum and growth factor in \emph{linear} perturbation theory), with its validity being confirmed by comparisons  to non-linear $N$-body simulations \citep[which has also allowed for the calibration of more accurate empirical fits, e.g.][]{2001MNRAS.321..372J, 2008ApJ...688..709T, 2016MNRAS.456.2486D}, there is no similar, simple approximation for \ac{FDM}.
None of the few existing (semi-)\linebreak[0]analytic estimates, such as using a Jeans-filtered power spectrum to account for the effects of \q{quantum pressure} \citep{2014MNRAS.437.2652M, 2017MNRAS.465..941D}, or a sharp $k$-space filtering in the Press–Schechter formalism with a variable cut-off \citep{2022MNRAS.510.1425K}, have been verified using the full non-linear \ac{SP} evolution, rendering their quantitative reliability still unclear.
The closest substitution of this goal thus far has been reached by comparing to the \ac{FDM} \ac{HMF} estimate in \citet{2016ApJ...818...89S}, which employed the approximate technique of collisionless $N$-body simulations with truncated initial fluctuation spectrum (but \emph{without} the full \ac{SP} dynamics), similar to those we have performed in this work as well ("\ac{CDM} with \ac{FDM} \acp{IC}").
This type of simulation technique is also how \ac{WDM} models are often computed \citep[e.\,g.][]{Lovell2014}, but importantly it exhibits a number of by now well-known difficulties \citep{2007MNRAS.380...93W}, such as the creation of \q{spurious} low-mass haloes that necessitate special removal procedures and ultimately introduce a significant source of uncertainty.

Supplementing our previous results \citep{2021MNRAS.506.2603M}, our new simulations allow for a measurement of the \ac{HMF} in a fully self-consistent cosmological wave simulation of \ac{FDM} for the first time.
For this purpose, we made use of the \ac{FoF}-based grid halo finder in {\AxiREPO} to identify collapsed structures in our \ac{SP} simulations.
However, as discussed in \cref{sec:halo-finding}, the cut-off in the \ac{FDM} initial power spectrum actually introduces considerable complications in identifying bound structures – some of which similarly plague \ac{WDM} simulations \citep{2007MNRAS.380...93W, 2013MNRAS.434.3337A} –, which mandate the application of additional filtering steps on the initial group catalogue determined by the halo finder. When implementing them, it turns out that only a limited number of most massive haloes in the raw catalogue survive in our \ac{FDM} simulation and can be considered physically robust, bound structures. Projections of these 68 haloes at $z = 3$ are displayed in \cref{fig:halo-projections}, making it clear that the majority of them exhibit a morphology in their outer parts that is quite distinct from what is typically seen in \ac{CDM} simulations. In particular, the haloes are usually embedded in thick filaments that show clear patterns of interference ridges.

In \Cref{fig:halo-mass-function} we show the resulting \ac{HMF} for the new \ac{FDM} simulations compared to the mass functions for \ac{CDM} \acp{IC}.
The results for the \ac{HMF} make it clear that the \acp{IC} have a much stronger impact than the choice of \ac{SP} or $N$-body dynamics.
Due to the cut-off on small scales in the initial \ac{FDM} power spectrum, the seeds of structure formation are suppressed, and it becomes virtually impossible for haloes below a certain mass threshold to form.
While the \ac{SP} dynamics also implies a threshold for the formation of virialised objects, this is drastically raised when introducing \ac{FDM} \acp{IC}.
Indeed, this behaviour is illustrated quite starkly by the fact that there are only 68 bound haloes in our entire simulation volume at $z = 3$, whereas the \ac{SP} simulation with \ac{CDM} \acp{IC} contains thousands of haloes in the same (quite moderate) volume at the same time. 

Although the statistics are somewhat poor due to the small number of resulting simulated haloes  –  despite our efforts to simulate a larger volume than has been standard for \ac{SP} simulations thus far  – we can measure the \ac{HMF} with reasonable accuracy and arrive at an important result:
There appears to be quite good agreement between the \ac{HMF} measured in our fully self-consistent \ac{FDM} wave simulation and the fitting function of \citet{2016ApJ...818...89S}, which was determined from $N$-body simulations with an \ac{FDM} power spectrum.
Consistent with our conclusion above, this confirms that the \acp{IC} are the primary factor in  determining the \ac{FDM} \ac{HMF}, and that the use of $N$-body simulations with filtering techniques is in principle sufficient to obtain at least an approximate estimate of the \ac{HMF} in an \ac{FDM} cosmology.

The (semi-)analytic estimates obtained in \citet{2017MNRAS.465..941D, 2022MNRAS.510.1425K}, on the other hand, differ from the result obtained using this technique:
Both feature a steeper turnover and drop at low masses than measured in simulations, and while the former results in a similar cut-off mass, it underestimates the number of haloes for high masses and overestimates it for low masses near the cut-off, whereas the latter features a lower cut-off mass, which does not account for the presence of the lower-mass haloes we find here.
Thus, compared to methods based on the extended Press–Schechter formalism, the $N$-body approximation seems to be the most reliable \q{simplified} approach to the \ac{HMF} so far, as it yields the best match to our measurements in full \ac{FDM} simulations.

Unfortunately, given the small number of haloes present even at our final redshift $z = 3$, it is not possible for us to meaningfully study the evolution of the \ac{HMF} with redshift in our simulation volume and to explicitly confirm that this conclusion also holds at other redshifts. Likewise, reliably characterising the quantitative precision of fitting functions such as those of \cite{2016ApJ...818...89S} will have to await  \ac{FDM} simulations with much larger volumes and hence better statistics.

\section{Fuzzy dark matter filaments}
\label{sec:filaments}

\begin{figure}
	\centering

%

	\includegraphics{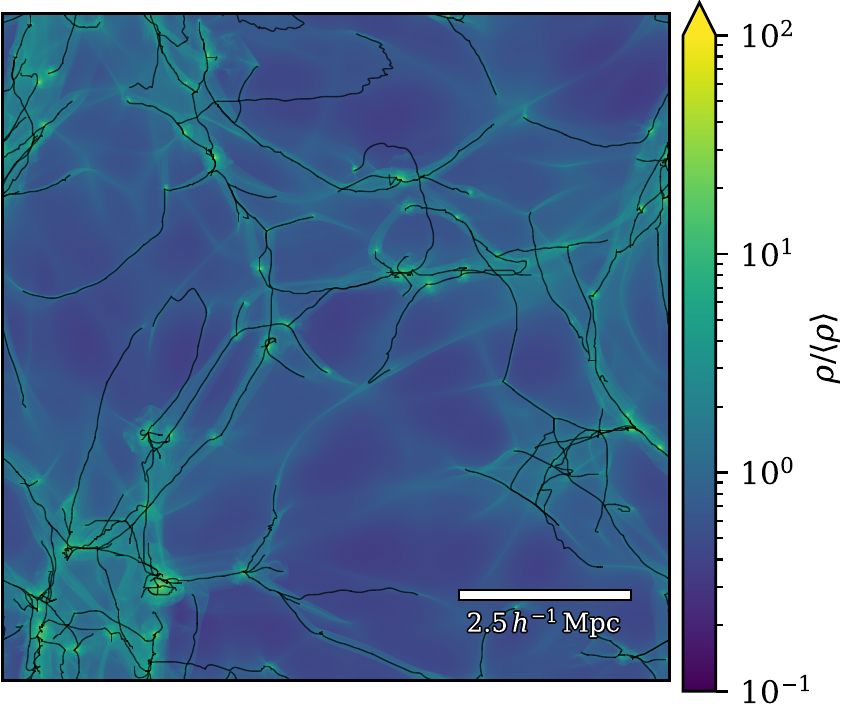}

	\includegraphics{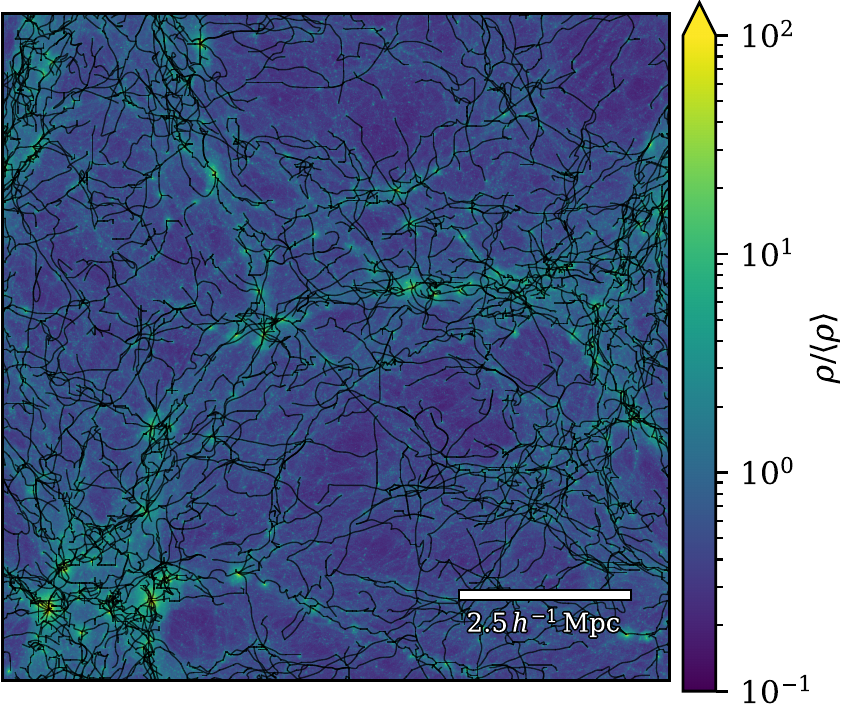}

	\caption{%
		Projected dark matter density showing the filaments identified by {\DisPerSE} (marked using black lines) in a cosmological box simulation of \acs*{FDM} at $z = 3$ with box size $L = \SI{10}{\per\hHubble\Mpc}$, grid size $N^3 = 8640^3$, \acs*{FDM} mass $m c^2 = \SI{7e-23}{\eV}$ (left), and in a similar \acs*{CDM} simulation with $2048^3$ particles (right).
		In order to enable {\DisPerSE} to process the simulated density field, it had to be scaled down to a $288^3$ (\ac{FDM}) or $260^3$ (\ac{CDM}) grid.%
	}
	\label{fig:mark-filaments}
\end{figure}

\begin{figure}
	\centering

	\includegraphics[scale=0.5]{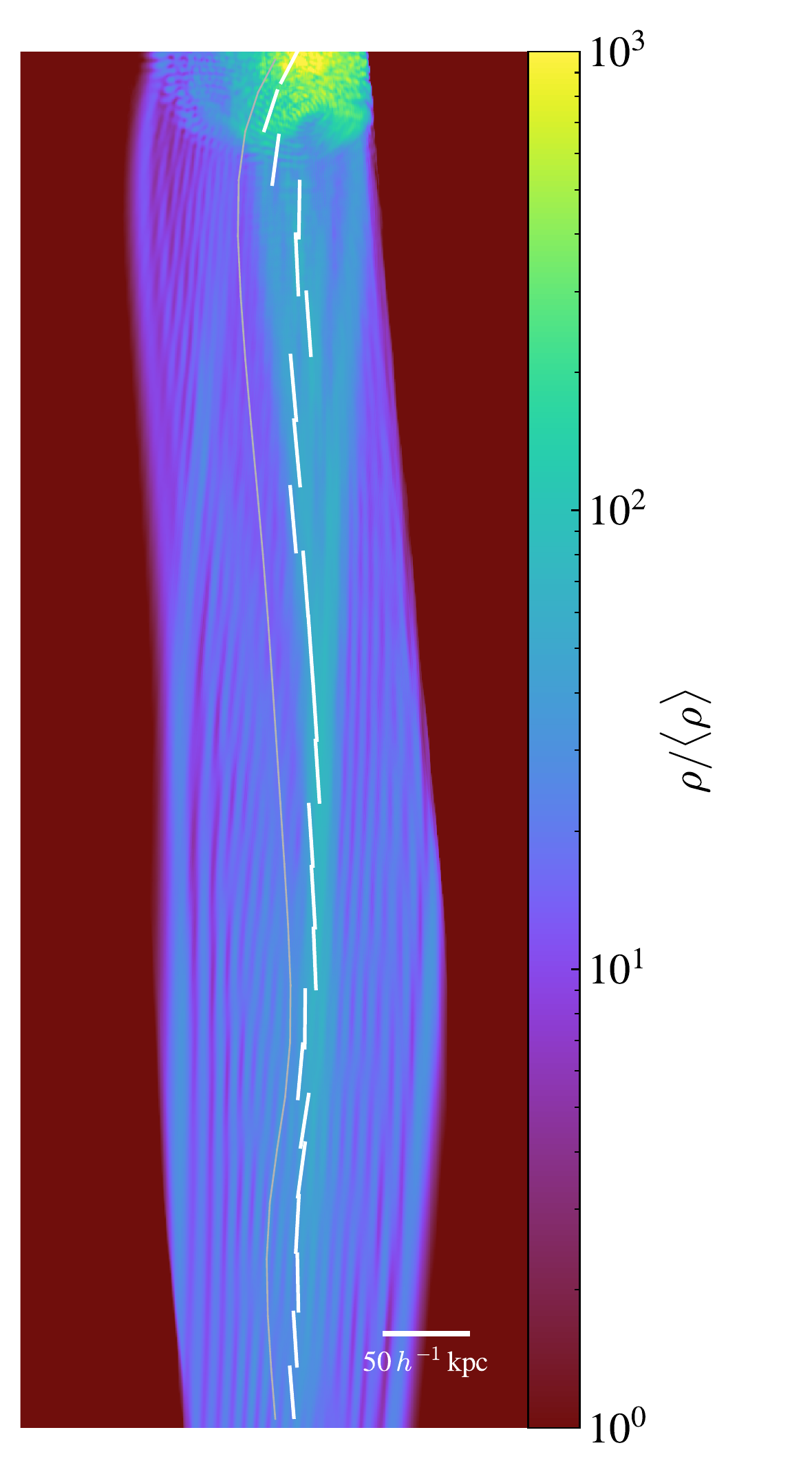}%

	\includegraphics[scale=0.5]{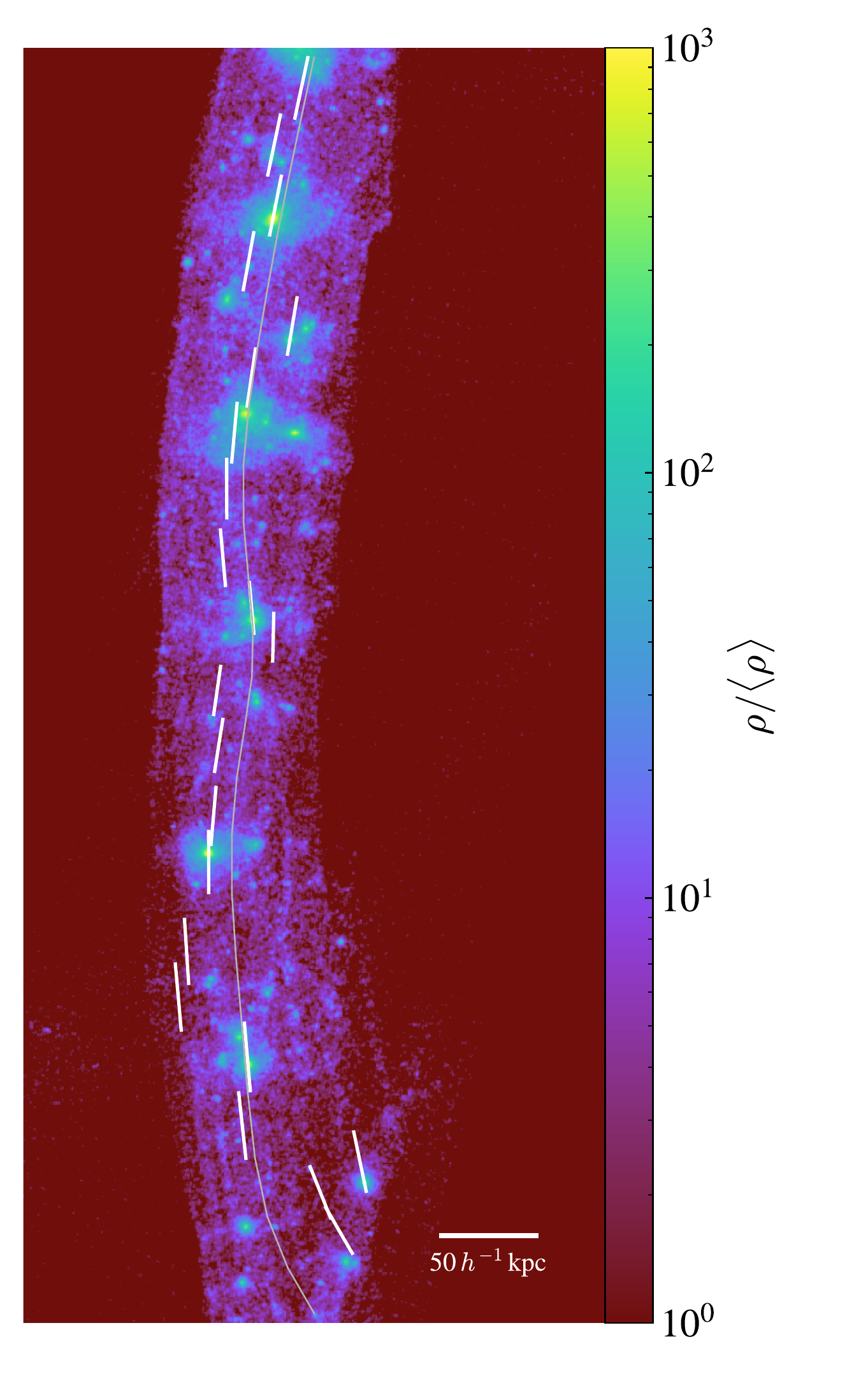}%

	\caption{%
		Projected dark matter density for an example \ac{FDM} (top) and \ac{CDM} (bottom) filament.
		A thin grey line indicates the location as identified by {\DisPerSE} on the coarse grid, while the thicker white line segments show how this initial {\DisPerSE} output was shifted to more accurately capture the filament centre for the purposes of computing radial density profiles.%
	}
	\label{fig:filament-examples}
\end{figure}

Our findings concerning the stark departure of the nature of filaments in \ac{FDM} compared to \ac{CDM} in \cref{sec:halo-finding}, and in particular their different visual appearance of the outer parts of haloes, compelled us to further investigate \ac{FDM} filaments.
In \cref{fig:mark-fof}, we provide an image of the cosmic web in \ac{FDM}, which is quite unlike the more familiar structures expected for \ac{CDM}, and is thus adding to this motivation.
Previous results for \ac{WDM} and smaller \q{proof-of-concept} simulations of \ac{FDM} have already provided hints that filaments represent another, second type of significant object when it comes to baryonic processes such as star formation \citep[e.\,g.][]{Gao2007, Gao2015, 2020MNRAS.494.2027M}.
Contrary to \ac{CDM}, where filamentary structures fragment into (sub-)haloes down to the smallest scales, the filaments in these models stay intact as smooth, overdense, large-scale dark matter structures.

We have used the {\DisPerSE} code \citep{2011MNRAS.414..350S, 2011MNRAS.414..384S} in order to attempt to identify the filaments in our \ac{FDM} simulations.
Unfortunately, this approach was met with technical difficulty due to computational limitations of the {\DisPerSE} software.
In particular, the code can only operate on shared-memory architectures, with no capabilities for today's distributed-memory parallel computing infrastructure using the \ac{MPI} standard.
This means that it is inherently limited to the amount of memory present on a single node on a computing cluster.
Gravely exacerbating the problem is the fact that the {\DisPerSE} algorithms are rather memory-intensive.
Even with the largest amount of memory available to us on the \ac{MPCDF}'s Raven system, which amounts to two terabytes, we were forced to severely \q{down-sample} our simulated density fields in order to allow them to be processed by {\DisPerSE}.
We note that previous work applying {\DisPerSE} to find filaments in large cosmological simulations, such as \citet{2020AA...641A.173G}, were able to avoid this problem by applying the code on a tracer field only, such as the distribution of galaxies.
However, this option is not available in our case, because there are no objects (such as haloes) present in the dark matter filaments which could serve as a tracer field with appropriate sampling rate. Rather we need to identify filaments directly in the smooth dark matter density field.

In order to ensure comparability between \ac{SP} (\ac{FDM}) and $N$-body (\ac{CDM}) simulations, we have constructed a \ac{CDM} density field by binning the $N$-body particles to a Cartesian grid using \ac{CiC} mass assignment, making the input to {\DisPerSE} a uniform Cartesian grid of the density field in both cases.
For \ac{FDM}, the original $8640^3$ grid was smoothed and scaled down by computing the means of neighbouring cells.
A cut value of $60ρ_{\mtext{m}}$ was used, which is the minimum threshold difference between critical points in order for {\DisPerSE} to keep them.
This ensures that the identified structural complex is not swamped by tiny small-scale fluctuations in the density field.

In the end, the largest grids which we could process with {\DisPerSE} were of size $288^3$ for our largest \ac{FDM} simulation, and $260^3$ for its \ac{CDM} counterpart.
This of course drastically reduces the spatial resolution (to \SI{34.7}{\per\hHubble\kpc} and \SI{38.5}{\per\hHubble\kpc}, respectively), and thus cannot be expected to yield either the full population of simulated filaments or their precise locations.
However, it nevertheless results in a reasonable filament catalogue which traces a good fraction of the cosmic web, as shown in \cref{fig:mark-filaments}.

While it is of course possible to investigate quantities like the radial filament density profiles even with this degraded resolution, the results are not very insightful since the identified filament lines will not generally trace the true, fully-resolved filament centres, making the computed profiles in
a significant fraction of the total radial extent unreliable due to the varying offsets from the centre.
In order to improve our measurement, we post-processed the filaments output by {\DisPerSE} by evaluating each filament segment and gradually shrinking a cylindrical region around the segment towards the centre of mass contained within the region at each step, similar to the "shrinking spheres" technique often used to determine the densest (central) point of a halo.
In this case, the segments were only allowed to be offset in a direction perpendicular to themselves, in order to avoid several segments clustering around a single dense point and preserving the property that the segments trace the full length of the filament.
While they do not remain continuously connected with this approach, the resulting error in the density profiles will be rather small as long as the originally identified filament line roughly traces the actual filament, since there will only be a slight inaccuracy in the measured distances of matter from the line centre due to the tilt of the segments compared to the "true" continuous filament line.

\begin{figure}
	\centering

	\import{res/}{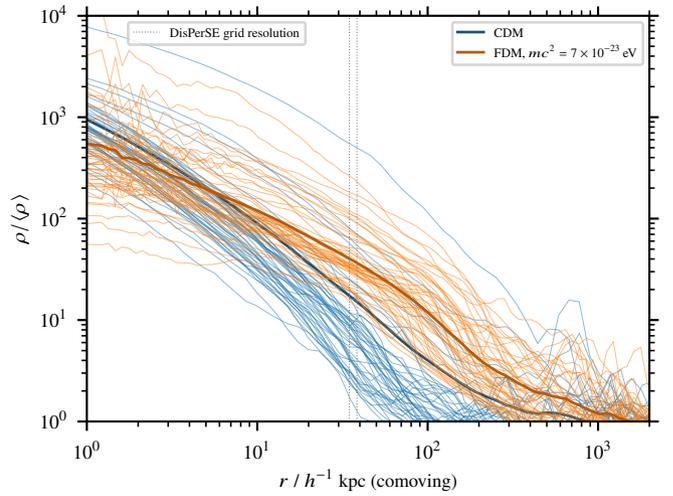}

	\caption{%
		Filament profiles at $z = 3$ for the filaments identified by {\DisPerSE} in $L = \SI{10}{\per\hHubble\Mpc}$ cosmological box simulations of \ac{CDM} ($2048^3$ particles) and \ac{FDM} ($N^3 = 8640^3$, $m c^2 = \SI{7e-23}{\eV}$), scaled down to $260^3$ and $288^3$ grids, respectively, for processing with {\DisPerSE}.
		The corresponding grid resolutions used with {\DisPerSE} are marked by dashed lines.
		Filaments identified by {\DisPerSE} were post-processed using the full simulation data, allowing to resolve the density profiles at much smaller distances than the resolution imposed by the {\DisPerSE} grids.
		The thin lines show a random sample of individual filament profiles, while the thick lines are the mean ("stacked") density profiles of all filaments with a total length $> \SI{350}{\per\hHubble\kpc}$.%
	}
	\label{fig:filament-profiles}
\end{figure}

Selected \ac{FDM} and \ac{CDM} filaments are shown in \cref{fig:filament-examples}, demonstrating the outcome of this procedure.
Especially for \ac{FDM}, it can clearly be seen that the original line which runs offset from the densest inner region of the filament is correctly shifted towards the centre.
For \ac{CDM}, on the other hand, the definition of a filament centre is more problematic, since the filaments are peppered with numerous small haloes, which are locally the densest points.
While the procedure also correctly shifts the segments to these dense points, the set of segments can become quite discontinuous in this case, and it is visually not clear that the haloes trace what one would intuitively identify as the centre of the filament as a whole.
It may be the case that another tracer, such as a suitably smoothed halo distribution, is more appropriate to describe the shape of \ac{CDM} filaments.

The final (cylindrically) radial density profiles, both as "stacked" profiles and for a sample of individual ones, can be found in \cref{fig:filament-profiles}.
It is evident that, for both \ac{FDM} and \ac{CDM}, filaments can reach rather high central densities.
The main difference between the two is the slope and extent of the density profiles:
The \ac{FDM} filament density remains relatively flat and falls more slowly towards the outer regions, reaching much further outwards before dropping to the background matter density, while \ac{CDM} filament profiles are relatively steep.
Whereas \ac{CDM} filaments on average decline around a radius of \SI{500}{\per\hHubble\kpc}, \ac{FDM} ones often extend beyond \SI{1}{\per\hHubble\Mpc}.

In the inner regions, \ac{CDM} filaments seem to reach higher central densities, although this statement should be interpreted with caution.
Firstly, as mentioned before, it should not be expected that the identified filaments represent the full population present in the simulation due to the coarse-graining which had to be performed on the density fields.
Secondly, the grid resolution for the \ac{FDM} simulation fundamentally limits the scales on which the density profile can be measured, whereas \ac{CDM} particles can gather on much smaller scales, so the innermost regions may be subject to resolution effects.
Thirdly, the filament segments tend to centre on haloes present in the \ac{CDM} filaments, which of course tend to have very high inner densities, meaning that the density profile measurement is in part "contaminated" by these halo profiles.
This once again raises the question of how to define the filament centre.
Finally, the scatter in \ac{FDM} filament densities is much greater than for \ac{CDM}, in particular featuring a population of filaments with inner densities of around $\num{e-2} ρ_{\mtext{m}}$ or even below.
Further investigation into different populations of filaments may reveal why these only seem to occur with \ac{FDM}.
Another interesting observation is that the inner \ac{FDM} filament profiles are much noisier than the \ac{CDM} ones.
While this can be partially attributed to resolution, which leads to poor sampling rates in the small inner cylindrical shells, it could also be a hint of the wave patterns and fluctuations present with \ac{FDM}.

Generally, it is clear that our procedure employed here is not the ideal way to measure dark matter filaments.
However, in the absence of superior methods and tools, \cref{fig:filament-profiles} demonstrates that it can still provide adequate results even far below the nominal grid resolution.
The main drawback of the approach is that the filament population, as identified by {\DisPerSE} on the coarse density grids, is likely incomplete, leading to potential biases in quantities like the mean density profiles.
Further, the sample is contaminated by pathological cases, such as spurious filaments which do not directly correspond to any structure in the original, fine density field.
Unfortunately, the need for "correcting" the output in post-processing makes it difficult to distinguish legitimate cases which can be fixed from pathological ones.
Enforcing a minimal filament length somewhat alleviates this, but is not a complete solution.

\section{Fuzzy dark matter halo profiles from a self-consistent cosmology}
\label{sec:halo-profiles}

\begin{figure}
	\centering

	\import{res/}{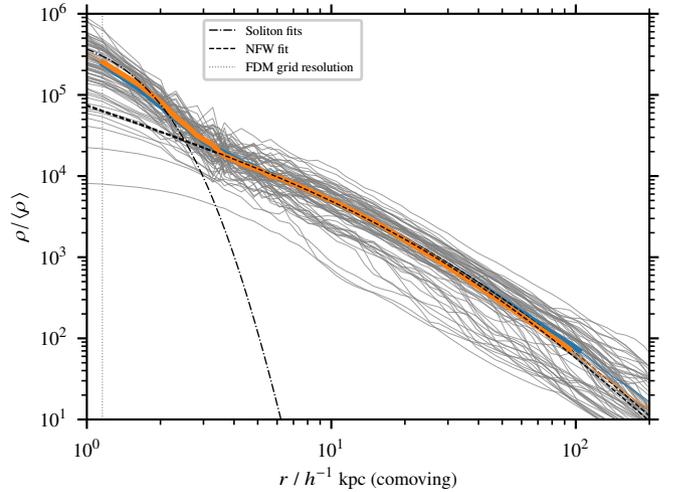}

	\caption{%
		Halo profiles for all 68 identified haloes (grey lines) in an $L = \SI{10}{\per\hHubble\Mpc}$ cosmological box simulation of \ac{FDM} ($N^3 = 8640^3$, $m c^2 = \SI{7e-23}{\eV}$, \ac{FDM} \acp{IC}).
		Additionally, the coloured lines show the \q{stacked} profile of all haloes for the \ac{FDM} \ac{IC} case and the subset of haloes which match these most closely in mass for the \ac{CDM} \ac{IC} case.
		The thin dashed line shows an \ac{NFW} fit (\cref{eq:nfw-fit}) to the region of the stacked profile within the virial radius $R_{200}$, which is indicated with a thick line.%
	}
	\label{fig:halo-profiles}
\end{figure}

The halo catalogues from our simulations in \citet{2021MNRAS.506.2603M} are also very useful to study a large sample of \ac{FDM} haloes, e.\,g.\ in the context of the core–halo mass relation \citep{2022MNRAS.511..943C}.
Due to the use of a \ac{CDM} initial matter power spectrum, the \ac{HMF} is strongly enhanced compared to a \q{pure} self-consistent \ac{FDM} model, even for low halo masses, as shown in \cref{sec:halo-mass-function}.
This significantly boosts the statistics by increasing the number of simulated haloes by several orders of magnitude, which can alleviate the issue of limited simulation volumes available in \ac{FDM} simulations to some extent.

However, there is the lingering question of whether the cosmological context, which differs from that of a \q{real} \ac{FDM} cosmology due to the use of a \ac{CDM} initial matter power spectrum, has an impact on the halo properties, and thus whether this approach is ultimately valid for studying halo profiles.  If this was the case, then the halo population may not be representative of a true sample of \ac{FDM} haloes with self-consistent \acp{IC}, making the approach less suitable for studying \ac{FDM} halo properties.  With the addition of \ac{FDM} \ac{IC} simulations of sufficient size, we are now able to explicitly check for differences in halo properties for the different \acp{IC}.

In \cref{fig:halo-profiles}, we show the radial density profile of all haloes in the largest \ac{FDM} simulation, as well as the \q{stacked} profile defined by these haloes.\footnote{%
	The procedure of \q{stacking} refers to taking the mean density across all haloes for each radial bin.%
}
In order to compare to the simulation with \ac{CDM} \acp{IC}, which contains many more haloes, we selected an equal number of haloes which most closely match those in the \ac{FDM} \ac{IC} simulation.
The resulting stacked profile is also plotted in \cref{fig:halo-profiles}, with both stacked profiles matching excellently – presenting a solitonic core in the centre and the familiar \ac{NFW} density profile \citep{1996ApJ...462..563N, Navarro1997},
\begin{equation}
	\label{eq:nfw-fit}
	ρ_{\mtext{c,NFW}}(r_{\mtext{c}}) =
	\frac{ρ_0}{\frac{r_{\mtext{c}}}{R_{\mtext{s}}} \ml(1 + \frac{r_{\mtext{c}}}{R_{\mtext{s}}}\mr)^2}
\end{equation}
in the outskirts.
Thus, while there might certainly be a statistical impact in the form that a larger sample of haloes is more likely to contain outliers, e.\,g.\ \q{unusual} haloes with a \q{rare} merger history, both cases agree very well at the level of the averaged density profiles, and our results do not show evidence for any systematic differences.

\section{Summary and Conclusions}
\label{sec:conclusions}

The simulations presented in this work constitute the largest numerical simulations of cosmic structure formation with \ac{FDM} that account for its full wave nature via the \ac{SP} equations, using a pseudo-spectral method.
By approaching simulation volumes more representative of cosmic large-scale structure, we have been able to follow the combined linear large-scale, and non-linear small-scale evolution of wave \ac{FDM} in a fully self-consistent manner.
The pseudo-spectral method can be considered the most accurate approach to solve the \ac{SP} system of equations numerically (although usually, the issue is not about the accuracy of the numerical approach to \ac{SP} equations, but rather whether an attempt is made to solve these equations at all, or whether the cost is shirked by choosing one of the approximate options).
It encompasses all aspects of the wave amplitude's temporal evolution, and in particular accounts for the oscillatory, order-unity fluctuations of the local density due to the wave effects in the axion-like dynamics.

We stress again that, unfortunately, the numerical resolution requirements to faithfully follow the \ac{FDM} dynamics are much more stringent than for the familiar $N$-body techniques applicable in the \ac{CDM} case.
Even large-scale modes require a very fine mesh due to the velocity criterion, \cref{eq:velocity-criterion-x}, since the spatial oscillations of the phase factors are otherwise not resolved, resulting in a halted evolution and eventually just numerical noise.
Nevertheless, the time step criterion, \cref{eq:time-step}, with its quadratic (rather than linear) dependence on the spatial resolution, enforces very small time steps when the mesh is made fine enough to resolve the de Broglie wavelength $λ_{\mtext{dB}}$, even when the coarsest possible mesh resolution is selected.
These requirements continue to severely limit the regimes available to full numerical simulations of \ac{FDM}, and in particular preclude large cosmological volumes to be computed at low resolution in a way similar to the standard practice in \ac{CDM}.

The large, fully self-consistent \ac{FDM} simulations we have performed here have allowed us to gain new insights into the evolution of the non-linear power spectrum in these cosmologies, especially in comparison to \ac{CDM} and the simplified approach of \q{\ac{CDM} with \ac{FDM} \acp{IC}}. We could also make the first direct, self-consistent measurements of the \acl{HMF} in such models.
In the process, we have discovered that the filamentary structure of the cosmic web behaves more similarly to \ac{WDM} than to \ac{CDM}, and we have characterised the properties of these filaments in comparison to \ac{CDM}.

Our main findings can be summarized as follows:
\begin{itemize}
\item Our study could
  determine the fully non-linear matter power spectrum of self-consistently evolved \ac{FDM} cosmologies over 
   an exceptionally broad range of scales, and we have also computed matching  \ac{CDM} N-body simulations with the same initial conditions.
   On large scales, we confirmed the consistency between all simulations – a manifestation of the Schrödinger–Poisson–Vlasov–Poisson correspondence, verifying the expectation that \ac{FDM} behaves like \ac{CDM} on these scales.
    On small scales, the \q{quantum pressure} in \ac{FDM} suppresses
    structure formation compared to \ac{CDM}. Self-consistently accounting for \ac{FDM} yields initial conditions with substantially reduced power on small scales, however, producing yet bigger suppression effects for small-small haloes, in a way strongly reminiscent of \ac{WDM} models.

\item We were able to draw a four-way comparison of the evolution of the matter power spectrum for the different choices of simulated dynamics and \acp{IC}.
	Our findings demonstrate that the differences between each option are time-dependent, with large variations in their relative behaviours.
	Especially at early times, the \acp{IC} are the determining factor, dominating the shape of the power spectrum at small scales due to the cut-off from at the \ac{FDM} Jeans scale.
	By $z = 15$, bottom-up structure formation is already in full swing for \ac{CDM}, with non-linear excess power accumulating on small scales.
	\ac{FDM} with \ac{CDM} \acp{IC} starts to reach the linear theory power spectrum again at this point, whereas simulations with \ac{FDM} \acp{IC} are still significantly lagging behind on scales smaller than the initial cut-off.
	Finally, for smaller redshifts, the shapes of all power spectra start to become rather similar, and by $z = 3$, the impact of going from \ac{CDM} to \ac{FDM} \acp{IC} and from $N$-body to \ac{SP} dynamics is roughly similar in size.
	Of course, only \ac{SP} simulations exhibit a small-scale excess \q{bump} (exceeding even the \ac{CDM} power) due to wave interference.

\item We have developed a technique which allows us to reliably identify haloes in self-consistent \ac{FDM} \ac{SP} simulations, including the proper \q{\ac{WDM}-like} \acp{IC} with suppressed small-scale power.
	This proved non-trivial because, similar to \ac{WDM}, standard halo finding algorithms like \ac{FoF} are not able to handle on their own the presence of extended, smoothly overdense filaments.
	By employing an increased overdensity threshold combined with a binding criterion and a filter on the identified objects' virial ratios, we have been able to identify the subset of gravitationally collapsed, virialised objects.

\item We could, for the first time, measure the self-consistent \acl{HMF} directly from pseudo-spectral \ac{FDM} simulations.
	By comparing \ac{SP} simulations with \ac{FDM} and \ac{CDM} \acp{IC}, we could confirm that the \ac{FDM} \acp{IC} have a larger impact on the \ac{HMF} than the late-time non-linear \ac{SP}  dynamics, although the suppression caused by the latter is still significant.
	For the \q{true} \ac{FDM} \ac{HMF}, we find broad agreement with the approximate scheme of obtaining the \ac{HMF} from $N$-body simulations with an \ac{FDM} power spectrum (\q{\ac{CDM} with \ac{FDM} \acp{IC}}).
	In contrast, predictions based on the extended Press–Schechter formalism \citep{2017MNRAS.465..941D, 2022MNRAS.510.1425K} do not fit our \ac{HMF} results so well.

\item We have investigated the differences between cosmic filaments for \ac{FDM} and \ac{CDM}.
	Like \ac{WDM}, \ac{FDM} filaments do not fragment into (sub-)haloes like those of \ac{CDM}, and instead feature the characteristic wave fluctuation patterns.
	We found significant differences in the shapes of filamentary density profiles, with \ac{FDM} filaments being much more extended and with a more slowly-declining density than \ac{CDM} ones.
	The scatter in identified filaments was also larger for \ac{FDM}, which features a population with very high densities (even comparable to lower-mass haloes).
	Such objects have been found to be the sites of first star formation with \ac{FDM} \citep{2020MNRAS.494.2027M}.

\item One advantage of the \q{\ac{FDM} with \ac{CDM} \acp{IC}} simulations which we have performed previously is the much larger population of simulated haloes, which can be used to infer \ac{FDM} halo properties using a large sample.
	However, it has not been a priori clear whether all the haloes obtained in such a simulation are representative of a fully self-consistent \ac{FDM} cosmology that is based on the correct primordial power spectrum, yielding only a much suppressed halo abundance. 
	In this work, we have been able to show that the properties of the simulated haloes match quite closely, validating the use of the larger halo catalogues for the purposes of studying \ac{FDM} haloes. This is reminiscent of $N$-body simulations of \ac{CDM} and \ac{WDM} cosmologies, which show very similar halo density profiles apart from a slightly reduced concentration of \ac{WDM} haloes, consistent with their later formation time \citep{Avila-Reese2001}.
\end{itemize}

Cosmological simulations of \ac{FDM} are numerically much more challenging than \ac{CDM}, but our work shows that at least small volumes can be studied with decent spatial resolution. Including baryons explicitly would clearly be a very worthwhile and interesting next step to  arrive at a more reliable assessment of whether these cosmologies are still viable, and how tight some of the constraints placed on the particle mass really are. In this regard, the high computational cost of \ac{FDM} simulations can actually be viewed as an encouragement. Unlike for \q{cheap} \ac{CDM} simulations, adding hydrodynamics and a modelling of galaxy formation physics should be comparatively easy to do – at least it is not expected to dominate the computational expense, and in this sense appears quite feasible. We intend to attempt this in forthcoming work.

\section*{Acknowledgements}

The authors would like to thank Daniela Galárraga-Espinosa for her help with the use of {\DisPerSE}, and Jens Stücker for a helpful discussion about methods for halo identification, as well as Mihir Kulkarni and Xiaolong Du for providing code to calculate the \ac{HMF} using their methods.
Computations were performed on the \ac{HPC} systems Cobra, Raven, and Freya at the \acf{MPCDF}.\footnote{%
	\url{https://www.mpcdf.mpg.de/}%
}

{\AxiREPO} (and {\AREPO}) make use of the \textcode{FFTW} \citep{Frigo2005}, \textcode{GSL} \citep{GSL}, and \textcode{HDF5} \citep{hdf5} libraries.
We employed the \textcode{GCC} \citep{GCC} and \textcode{Open MPI} \citep{gabriel04:_open_mpi} implementations of the \textcode{C} programming language and the \textcode{MPI} standard.
The following software and libraries were additionally used for data analysis and production of figures:
\textcode{Python} \citep{Python-CWI,Python}, \textcode{NumPy} \citep{harris2020array}, \textcode{SciPy} \citep{2020SciPy-NMeth}, \textcode{Numba} \citep{10.1145/2833157.2833162}, \textcode{Matplotlib} \citep{Hunter:2007}, \textcode{mpi4py} \citep{DALCIN20051108,9439927}, \textcode{Astropy} \citep{2013ascl.soft04002G,2018AJ....156..123A}, \textcode{yt} \citep{2010ascl.soft11022T,2011ApJS..192....9T}.

\section*{Data availability}

The data underlying this article will be shared upon reasonable request to the corresponding author.

\renewcommand*{\refname}{REFERENCES}
\bibliographystyle{mnras}
\bibliography{bib}

\newpage
\appendix

\section{Small inconsistency in the fuzzy dark matter initial conditions}

\begin{figure}
	\includegraphics[width=\columnwidth]{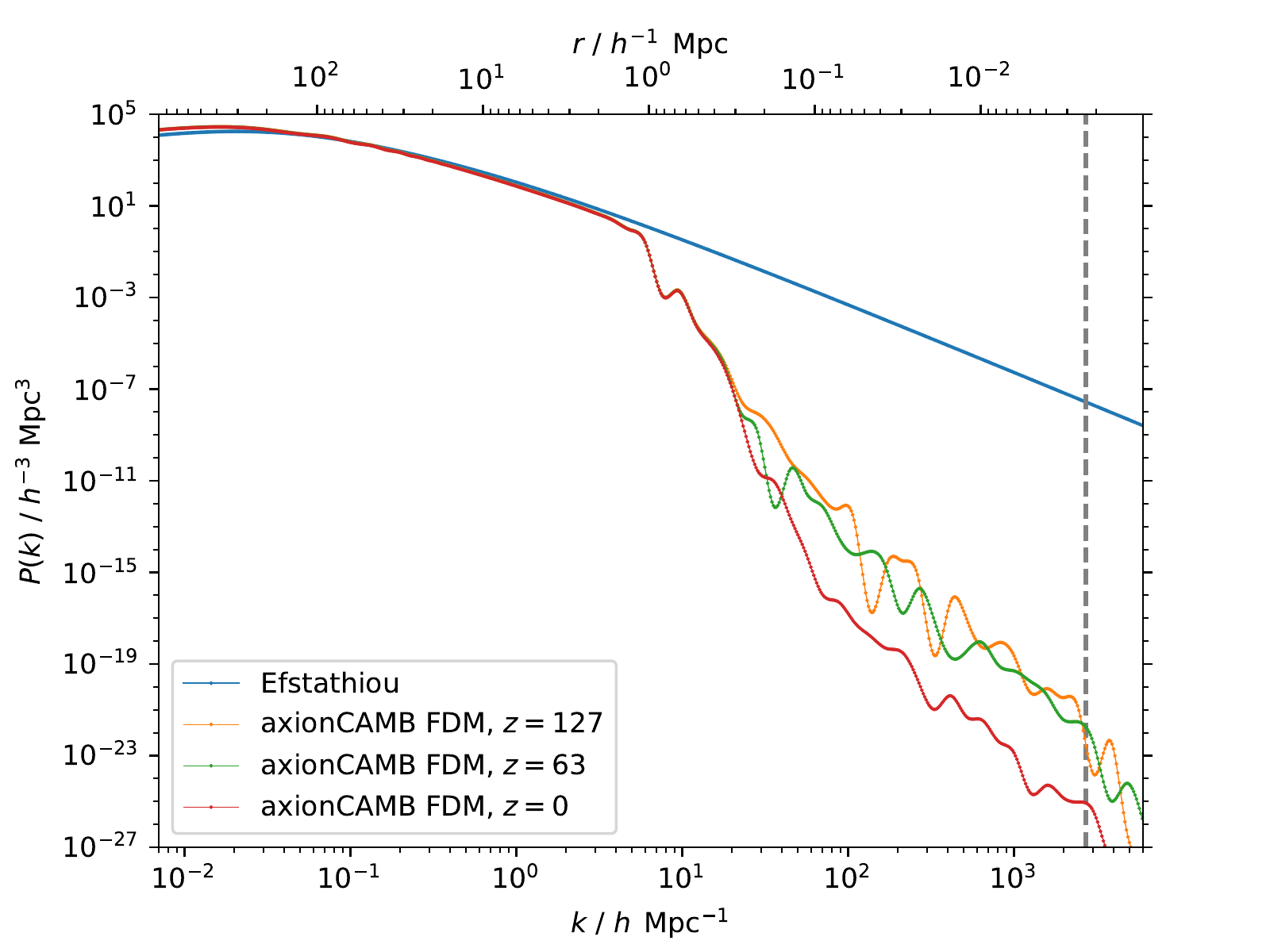}
	\caption{%
		Linear \ac{FDM} power spectra ($m c^2 = \SI{7e-23}{\eV}$) generated by {\axionCAMB} for different redshifts $z$, with the \citet{1990Natur.348..705E,1992MNRAS.258P...1E} power spectrum from \cref{eq:efstathiou} for comparison.
		For easier comparison, all power spectra were scaled to $z = 0$ using the linear \ac{CDM} growth factor.
		The dashed vertical line indicates the grid resolution scale for a simulation in a \SI{10}{\Mpc\per\hHubble} box with an $8640^3$ grid.%
	}
	\label{fig:axioncamb-redshifts}
\end{figure}

After performing the simulations with \ac{FDM} \acp{IC}, we discovered that, unfortunately, a small error had occurred with the used input power spectra (generated by \axionCAMB):
Instead of generating the power spectrum at the correct starting redshift of $z = 127$, the power spectrum was generated for $z = 0$ and then scaled back to $z = 127$ using the (scale-independent) \emph{\ac{CDM}} growth factor.
However, since the growth factor for \ac{FDM} is scale-dependent, procedure does not result in the correct power spectrum for an earlier redshift.

\Cref{fig:axioncamb-redshifts} shows how the shape of the linear \ac{FDM} power spectrum compares for different redshifts.
In this case, the red line ($z = 0$) was used (incorrectly) instead of the orange line ($z = 127$).
Fortunately, the discrepancy is not very serious.
Since \ac{FDM} coincides with \ac{CDM} on large scales, there is no error incurred there.
The difference essentially boils down to a slightly increased suppression on scales smaller than \SI{300}{\per\hHubble\kpc}, accompanied by some changes in the details of the oscillations of the power spectrum on these scales.
However, the order of magnitudes of the relative suppression involved (compared to \ac{CDM}) are \num{e-12} and \num{e-9}, or in other words, structure is effectively completely suppressed on these scales in both cases.
For all intents and purposes, the difference is negligible, and is erased by the vastly larger amount of power coming in from non-linear power transfer in the later stages of evolution (cf.\ \cref{sec:power-spectrum}).
Nevertheless, we would like to note this slight inconsistency in the \acp{IC} here for completeness.

\bsp
\label{lastpage}
\end{document}